\documentclass[a4paper,twocolumn,11pt,accepted=2021-03-10]{quantumarticle}
\pdfoutput=1
\usepackage[utf8]{inputenc}
\usepackage[pdftex]{graphicx}
\usepackage{float}
\usepackage{amsfonts,amsmath,amssymb,amsthm}
\usepackage{stmaryrd}
\usepackage{dsfont}
\usepackage{array}
\usepackage{bm}
\usepackage{subcaption}
\usepackage{mathrsfs}
\usepackage{pifont}
\usepackage{multirow}
\usepackage{upgreek}
\usepackage{xcolor}
\usepackage[pdftex,
            pdftitle={Dimensional Expressivity Analysis of Quantum Circuits},
            pdfauthor={Lena Funcke,Tobias Hartung,Karl Jansen,Stefan Kuehn,Paolo Stornati},
            bookmarks,
            colorlinks,            
            menucolor=black,            
            plainpages=false,
            linkcolor=myblue,
            citecolor=mymagenta,
            menucolor=black,
            urlcolor=myblue,
            pdfpagelabels,
            hypertexnames=false]{hyperref}
\usepackage{verbatim}
\usepackage{slashed}
\usepackage[ruled,vlined]{algorithm2e}
\usepackage[braket,qm]{qcircuit}
\usepackage[numbers,sort&compress]{natbib}
\definecolor{mymagenta}{RGB}{200, 0, 100}
\definecolor{myblue}{RGB}{45, 48, 146}

\newcommand{\cn}{\ensuremath{\mathbb{C}}}
\newcommand{\rn}{\ensuremath{\mathbb{R}}}
\newcommand{\zn}{\ensuremath{\mathbb{Z}}}
\newcommand{\nn}{\ensuremath{\mathbb{N}}}
\newcommand{\Hp}{\ensuremath{\mathcal{H}}}
\newcommand{\fa}{\forall}
\newcommand{\ex}{\exists}
\newcommand{\sse}{\ensuremath{\subseteq}}

\renewcommand{\d}{\ensuremath{\partial}}
\renewcommand{\l}{\ensuremath{\left}}
\renewcommand{\r}{\ensuremath{\right}}
\newcommand{\lnorm}{\left\lVert}
\newcommand{\rnorm}{\right\lVert}
\newcommand{\lbetr}{\left\lvert}
\newcommand{\rbetr}{\right\lvert}
\newcommand{\norm}[1]{\lnorm {#1}\rnorm}

\newcommand{\abs}[1]{\lbetr {#1}\rbetr}

\newcommand{\ord}{\ensuremath{\mathrm{ord}}}
\newcommand{\Eig}{\ensuremath{\mathrm{Eig}}}
\newcommand{\rank}{\ensuremath{\mathrm{rank}}}
\newcommand{\id}{\ensuremath{\mathrm{id}}}
\newcommand{\dist}{\ensuremath{\mathrm{dist}}}
\newcommand{\const}{\ensuremath{\mathrm{const.}}}
\newcommand{\CNOT}{\ensuremath{\mathrm{CNOT}}}
\newcommand{\SWAP}{\ensuremath{\mathrm{SWAP}}}
\newcommand{\anc}{\ensuremath{\mathrm{anc}}}
\newcommand{\prob}{\ensuremath{\mathrm{p}}}

\renewcommand{\phi}{\varphi}

\renewcommand{\rho}{\varrho}

\renewcommand{\theta}{\vartheta}
\newcommand{\eps}{\varepsilon}
\DeclareMathAlphabet{\mathesstixfrak}{U}{esstixfrak}{m}{n}
\graphicspath{{pics/}}

\newtheorem{theorem}{Theorem}[section]
\newenvironment{example*}[1][Example]{\begin{trivlist}
  \item[\hskip \labelsep {\rmfamily\scshape #1}]}{\nopagebreak\flushright$\square$\end{trivlist}}

\newcommand{\aref}[1]{\hyperref[#1]{App.~\ref{#1}}}

\begin{document}

\title{Dimensional Expressivity Analysis of Parametric Quantum Circuits}
 \author{Lena Funcke}
 \affiliation{Perimeter Institute for Theoretical Physics, 31 Caroline Street North, Waterloo, ON N2L 2Y5, Canada}
 \orcid{0000-0001-5022-9506}
 \author{Tobias Hartung}
 \affiliation{Computation-Based  Science  and  Technology  Research  Center, The  Cyprus  Institute,  20  Kavafi  Street,  2121  Nicosia,  Cyprus}
 \affiliation{Department of Mathematics, King’s College London, Strand, London WC2R 2LS, United Kingdom}
 \orcid{0000-0001-6133-5232}
 \author{Karl Jansen}
 \affiliation{NIC, DESY Zeuthen, Platanenallee 6, 15738 Zeuthen, Germany}
 \orcid{0000-0002-1574-7591}
 \author{Stefan K{\"u}hn}
 \affiliation{Computation-Based  Science  and  Technology  Research  Center, The  Cyprus  Institute,  20  Kavafi  Street,  2121  Nicosia,  Cyprus}
 \orcid{0000-0001-7693-350X}
 \author{Paolo Stornati}
 \affiliation{NIC, DESY Zeuthen, Platanenallee 6, 15738 Zeuthen, Germany}
 \affiliation{Institut für Physik, Humboldt-Universität zu Berlin, Zum Großen Windkanal 6, D-12489 Berlin, Germany}
 \orcid{0000-0003-4708-9340}
\date{(Dated: November 21, 2020)}

\begin{abstract}
  Parametric quantum circuits play a crucial role in the performance of many variational quantum algorithms. To successfully implement such algorithms, one must design efficient quantum circuits that sufficiently approximate the solution space while maintaining a low parameter count and circuit depth. In this paper, we develop a method to analyze the dimensional expressivity of parametric quantum circuits. Our technique allows for identifying superfluous parameters in the circuit layout and for obtaining a maximally expressive ansatz with a minimum number of parameters. Using a hybrid quantum-classical approach, we show how to efficiently implement the expressivity analysis using quantum hardware, and we provide a proof of principle demonstration of this procedure on IBM's quantum hardware. We also discuss the effect of symmetries and demonstrate how to incorporate or remove symmetries from the parametrized ansatz.
\end{abstract}

\maketitle

\section{Introduction}

Current noisy intermediate-scale quantum (NISQ) computers~\cite{Preskill2018} open up a new route to tackle a variety of computational problems that cannot be addressed efficiently with classical computers. Potential applications range from machine learning~\cite{Biamonte2017} to finance~\cite{Orus2019} to various optimization problems~\cite{Montanaro2016,Brandao2017}. In particular, quantum computers intrinsically evade the sign problem that hinders Monte Carlo simulations of strongly correlated quantum-many body problems in certain parameter regimes~\cite{Troyer2005}. While current hardware is still of limited size and suffers from a considerable level of noise, it has already been successfully demonstrated that NISQ devices can outperform classical computers~\cite{Arute2019}. Moreover, techniques for mitigating the effects of noise are rapidly developing~\cite{Temme2017,Kandala2017,Endo2018,YeterAydeniz2020,Funcke2020}.

Parametrized quantum circuits are at the heart of many algorithms designed for NISQ devices. A prominent example is variational quantum simulation (VQS)~\cite{Peruzzo2014,McClean2016}, a class of hybrid quantum-classical algorithms for solving optimization problems. Using a quantum circuit composed of parametric gates as an ansatz, the quantum coprocessor is used to efficiently evaluate the cost function for a given set of variational parameters. The minimization procedure is then performed on a classical computer in a feedback loop based on the measurement outcome obtained from the quantum device. VQS has been successfully applied to quantum many-body systems in quantum chemistry~\cite{Peruzzo2014,Wang2015,Kandala2017,Hempel2018} and even quantum mechanics and quantum field theory~\cite{Kokail2018,Hartung2018,Jansen2019,Paulson2020,Haase2020}.

The design of the parametrized quantum circuit used as an ansatz is crucial for the success of a VQS approach. While it should be sufficiently expressive to be able to approximate the solution of the problem to good accuracy, NISQ devices only allow for executing circuits of short depths faithfully before noise starts to dominate. Thus, one should avoid superfluous parameters in the circuit resulting in unnecessary gate operations. At the same time, having an insufficient number of independent parameters in the circuit can lead the classical minimization algorithm to fall into false local minima. A careful design of the quantum circuit is therefore essential to make optimal use of current NISQ devices.

This question of circuit expressivity is an active area of research, see Refs.~\cite{Geller_2018,Sim2019,Bataille2020,Sim2020,Rasmussen2020,Hubregtsen2020,Schuld2020,Fontana2020, Gard_2020,Barron2020,Kim2020} and references therein. In particular, it has been proposed to assess the expressivity of a parametric quantum circuit by quantifying the circuit's ability to uniformly reach the full Hilbert space~\cite{Sim2019}, which was accomplished by computing statistical properties based on randomly sampling states from a given circuit template. Using this ansatz, it has been shown that the number of single-qubit rotations in certain parameterized quantum circuits can be decreased without compromising the expressivity of the circuit~\cite{Rasmussen2020}. On the other hand, it has been shown that the inclusion of redundant parametrized gates can make the quantum circuits more resilient to noise~\cite{Fontana2020,Kim2020}. In particular, Ref.~\cite{Kim2020} showed that increasing the depth of the circuit above a certain threshold, which heavily overparameterizes the problem, leads to a good approximation of the ground state assuming a noise-free quantum device. Thus, in practice, these two effects of increasing the circuit depth and the resilience to noise have to be weighted against each other, especially for different amounts of noise in the given circuit. A practical approach to circuit design should therefore incorporate both aspects. For example, the here proposed dimensional expressivity analysis can be used to ensure sufficient expressivity of a minimal circuit which could then be overparametrized to harness the positive effects studied in Ref.~\cite{Kim2020}.

For non-parametric gates, it is also possible to follow an algebraic approach, as demonstrated in Ref.~\cite{Bataille2020} for the group generated by $\CNOT$ and $\SWAP$ gates. The reduction rules developed in Ref.~\cite{Bataille2020} can then be employed to minimize the number of $\CNOT$ and $\SWAP$ operations in a given quantum circuit, while the action of the quantum circuit, i.e., the input-to-output relation, is unchanged. 

In this paper, we will devise a general method to analyze the expressivity of a given parametric quantum circuit and to identify redundant parameters, which is independent of the statistical approach proposed in Ref.~\cite{Sim2019}. In fact, we will develop a geometric approach to minimize the number of parameters in a parametric circuit, while the set of states that can be generated by the circuit remains unchanged. In this sense, the analysis proposed in this paper complements the algebraic approach~\cite{Bataille2020}, which can minimize the number of non-parametric gates, while our proposed approach minimizes the number of parametric gates.

To this end, we consider quantum circuits as operators that act on a state within a certain Hilbert space, typically the state a given quantum device is initialized in. Thus, we can consider parametric quantum circuits as a map from the set of parameters to a subset of the state space of the quantum device. In many cases, this image set of reachable states by the quantum circuit forms a submanifold of the state space of the quantum device. As such, the (dimensional) expressivity of the quantum circuit can be understood as the dimension of this manifold of reachable states. We will also call this manifold the circuit manifold.

In the ideal case, the dimension of the circuit manifold coincides with the number of parameters of the quantum circuit. This implies that there are no redundant parameters. Similarly, in an ideal situation, the circuit should be able to reach all physical states. The set of physical states of course depends on the model simulated using the VQS and can be much more restrictive than the full state space of the quantum device. For example, if the physical model has a certain symmetry, e.g., $\zn_2$ symmetry for fermionic models or translational invariance in homogeneous materials, then any state in the state space of the quantum device that does not satisfy these symmetries would be considered unphysical. For efficient quantum simulations of such models, the quantum circuit should therefore be able to reach all physical states (to be maximally expressive) while simultaneously not generate unphysical states.

The main goal of efficient circuit design is therefore to generate all physical states without introducing unphysical states or redundant parameters. In most cases, designing a circuit that satisfies physical symmetries is much easier than ensuring maximal expressivity while avoiding redundancy. Both maximal expressivity and avoiding redundancy can locally be tested using dimensional expressivity analysis. If there are more parameters than the circuit manifold dimension, then superfluous parameters exist and can be removed using the dimensional analysis techniques developed in this paper. 

Similarly, the codimension\footnote{The codimension of a sub-manifold $M'$ of a finite-dimensional manifold $M$ is defined as the difference between the dimensions: $\text{codim}(M')=\dim(M)-\dim(M')$.} of the circuit manifold with respect to the manifold of physical states provides a measure of expressivity deficiency. If the codimension vanishes, then both manifolds coincide (locally), which ensures maximal expressivity of the quantum circuit. This is because the circuit manifold must be a sub-manifold of the manifold of physical states (provided all states generated by the quantum circuit are physical) and having the same dimension thus locally implies that they coincide. Additionally, the codimension, as a measure of expressivity deficiency, is a predictor of the number of parameters missing for a maximally expressive circuit. Thus, the dimensional expressivity analysis developed in this paper will provide information that can guide the construction of gates to be added to the quantum circuit. 

The dimensional expressivity analysis developed in this paper will (i)~allow us to determine which parameters are superfluous in the quantum circuit and (ii)~tell us how to modify a given quantum circuit to eliminate these parameters, while respecting
the symmetries of the physical model. In addition, we demonstrate how to modify a given quantum circuit to remove unwanted symmetries from the states generated by the quantum circuit, and how to approach the construction of a maximally expressive quantum circuit respecting the physical symmetries. As an example of a common symmetry found in physical models, we discuss translational invariance and describe how the physical state space is divided into sectors defined by the eigenvalues of the momentum operator.

The rest of the paper is structured as follows. In \autoref{sec:EfficientSU2}, we provide a simple introductory example of how to analyze the dimensional expressivity of a quantum circuit. We chose to analyze QISKIT's \verb|EfficientSU2| 2-local circuit as an example of a commonly chosen circuit layout.

After this initial example, we then provide the general mathematical setup and develop the dimensional expressivity analysis in \autoref{sec:parameter_reduction}. In particular, \autoref{sec:EfficientSU2_again} contains the technical details for QISKIT's \verb|EfficientSU2| 2-local circuit as discussed in \autoref{sec:EfficientSU2}. Although \autoref{sec:parameter_reduction} is mathematically very technical, we embed further illustrative examples for the mathematical statements obtained.

In \autoref{sec:4Q_circuit_design}, we apply the dimensional expressivity analysis
to the case of a translationally invariant 4-qubit circuit. This circuit was adapted from a well-working non-translationally invariant quantum circuit. However, incorporation of translational invariance led to a number of problems, such that the VQS algorithm failed to find the ground state. The example discussed in \autoref{sec:4Q_circuit_design} therefore highlights how to use the dimensional expressivity analysis to identify circuit design shortcomings and how to use the obtained information to custom design a maximally expressive circuit.

In \autoref{sec:translation_invariance}, we generalize the ideas underlying the example of \autoref{sec:4Q_circuit_design} and analyze translationally invariant quantum circuits and momentum sectors of the physical state space. This serves as an example analysis for any type of physical symmetry that is generated through spectral invariants. As such, the ideas in \autoref{sec:translation_invariance} can be generalized further to obtain dimensions of physical state spaces and the necessary information underlying the custom design of maximally expressive quantum circuits.

Reducing the complete state space of a quantum device to the physical state space of a given model furthermore introduces questions surrounding the ability of a VQS to find the solution. For VQSs using a local optimizer in parameter space, this effectively translates into the question whether or not the physical state space is path connected and whether or not a path from the initial parameter set to a parameter set for the solution can be found using only information available in the physical state space, that is, information that can be obtained from quantum device calls. These path connectedness questions are addressed in \aref{app:eigenspaces_path_connected}, since they are highly technical and explore a direction that is not necessary to delve into deeply for the examples provided in this paper. However, they can be helpful in applications of the dimensional expressivity analysis because they address some non-trivial mathematical properties of the dimensional expressivity analysis that can lead to unexpected behaviors such as failure of the VQS to converge.

Since the dimensional expressivity analysis requires complete state information, it is unlikely that an efficient classical algorithm exists to perform this analysis. In particular, any classical algorithm is likely to require exponential memory in the number of qubits even if there are only two parameters in the given circuit. In \autoref{sec:implementation}, we will therefore discuss a hybrid quantum-classical algorithm performing the dimensional expressivity analysis. Under some assumptions on the type of parametric gates used, this algorithm performs at polynomial cost in the number of circuit parameters, for both the classical part of the algorithm and in the number of quantum device calls. The memory requirements are polynomial in the number of circuit parameters for the classical part of the dimensional expressivity analysis, while the quantum part of the algorithm requires one ancilla qubit and six additional gates (independent of the number of qubits and parameters, cf. \autoref{sec:hybrid-algorithm}). As such, the proposed dimensional expressivity analysis has the potential to be automated and could be used for on the fly circuit optimization. A hardware implementation on IBMQ  is discussed in \autoref{sec:hardware-implementation}. The technical details of the hardware experiments are presented in \aref{app:hardware_implementation}.

We summarize and discuss our results in \autoref{sec:conclusion}.

\section{QISKIT's Efficient SU(2) 2-local circuit\label{sec:EfficientSU2}}
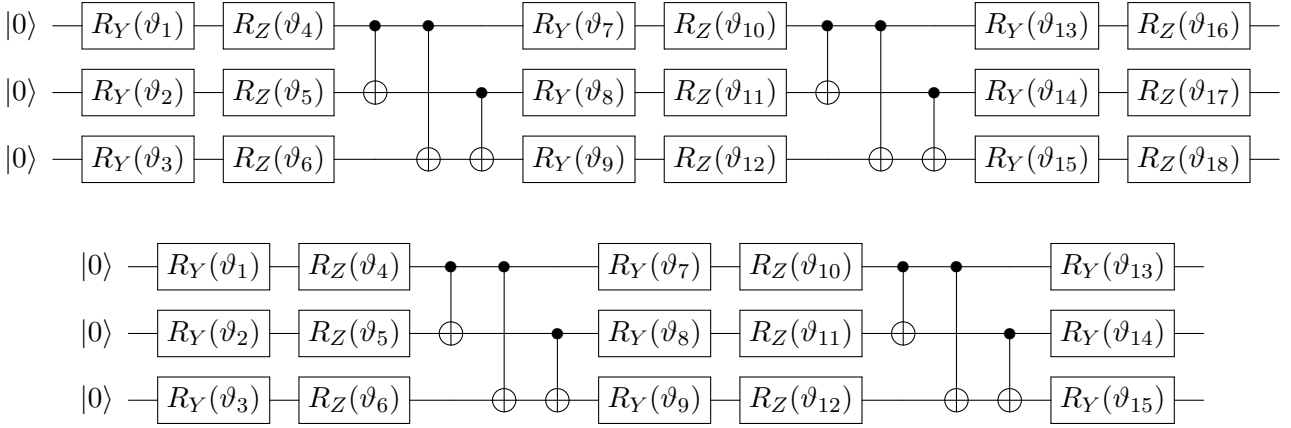
\begin{figure*}[!htp]
  \begin{align*}
    \Qcircuit @C=1em @R=.7em {
      \lstick{\ket{0}} & \gate{R_Y(\theta_1)} & \gate{R_Z(\theta_4)} & \ctrl{1} & \ctrl{2} & \qw      & \gate{R_Y(\theta_7)} & \gate{R_Z(\theta_{10})} & \ctrl{1} & \ctrl{2} & \qw      & \gate{R_Y(\theta_{13})} & \gate{R_Z(\theta_{16})} & \qw \\
      \lstick{\ket{0}} & \gate{R_Y(\theta_2)} & \gate{R_Z(\theta_5)} & \targ    & \qw      & \ctrl{1} & \gate{R_Y(\theta_8)} & \gate{R_Z(\theta_{11})} & \targ    & \qw      & \ctrl{1} & \gate{R_Y(\theta_{14})} & \gate{R_Z(\theta_{17})} & \qw\\
      \lstick{\ket{0}} & \gate{R_Y(\theta_3)} & \gate{R_Z(\theta_6)} & \qw      & \targ    & \targ    & \gate{R_Y(\theta_9)} & \gate{R_Z(\theta_{12})} & \qw      & \targ    & \targ    & \gate{R_Y(\theta_{15})} & \gate{R_Z(\theta_{18})} & \qw
    }
  \end{align*}  
  \begin{align*}
    \Qcircuit @C=1em @R=.7em {
      \lstick{\ket{0}} & \gate{R_Y(\theta_1)} & \gate{R_Z(\theta_4)} & \ctrl{1} & \ctrl{2} & \qw      & \gate{R_Y(\theta_7)} & \gate{R_Z(\theta_{10})} & \ctrl{1} & \ctrl{2} & \qw      & \gate{R_Y(\theta_{13})} & \qw \\
      \lstick{\ket{0}} & \gate{R_Y(\theta_2)} & \gate{R_Z(\theta_5)} & \targ    & \qw      & \ctrl{1} & \gate{R_Y(\theta_8)} & \gate{R_Z(\theta_{11})} & \targ    & \qw      & \ctrl{1} & \gate{R_Y(\theta_{14})} & \qw\\
      \lstick{\ket{0}} & \gate{R_Y(\theta_3)} & \gate{R_Z(\theta_6)} & \qw      & \targ    & \targ    & \gate{R_Y(\theta_9)} & \gate{R_Z(\theta_{12})} & \qw      & \targ    & \targ    & \gate{R_Y(\theta_{15})} & \qw
    }
  \end{align*}  
  \caption{Upper panel: QISKIT's \texttt{EfficientSU2} 2-local circuit with $N=2$ for three qubits. Lower panel: Reduction of QISKIT's \texttt{EfficientSU2} 2-local circuit with $N\ge2$ to a maximally expressive circuit for three qubits using the minimal number of parameters.}
  \label{fig:EffSU2}
\end{figure*}

As a simple introductory example, let us analyze QISKIT's~\cite{Abraham2019} hardware efficient SU(2) 2-local circuit \verb|EfficientSU2(3, reps=N)|, which the QISKIT documentation~\cite{QISKIT:EfficientSU2} proposes as ``a heuristic pattern that can be used to prepare trial wave functions for variational quantum algorithms or classification circuit for machine learning.'' This circuit consists of $N+1$ blocks of $R_Y$ and $R_Z$ gates applied to every qubit. These blocks are interlaced with $N$ blocks containing $\mathrm{CNOT}(q,q')$ gates for all $q<q'$. For example, setting $N=2$, this circuit is given in the upper panel of \autoref{fig:EffSU2} on the next page. 

The method allows the user to input various parameters in order to change the layout of this circuit, including different gates in the $N+1$ single-qubit gate blocks and different entanglers, such as a circular $\mathrm{CNOT}$ pattern $\mathrm{CNOT}(q,q+1)$, i.e., 
\begin{align*}
  \Qcircuit @C=1em @R=.7em {
    \lstick{} & \targ     & \ctrl{1} & \qw      & \qw      & \qw\\
    \lstick{} & \qw       & \targ    & \ctrl{1} & \qw      & \qw\\
    \lstick{} & \qw       & \qw      & \targ    & \ctrl{1} & \qw\\
    \lstick{} & \ctrl{-3} & \qw      & \qw      & \targ    & \qw  
  }
\end{align*}
in the four-qubit case (to make the pattern more explicit). Leaving all optional parameters at their default value yields the circuit in the upper panel of \autoref{fig:EffSU2}.

Our aim for this section is to highlight some of the knowledge that can be gained using the dimensional expressivity analysis for the \verb|EfficientSU2(3, reps=N)| circuit. As such, the analysis will be non-technical. For technical details we refer to \autoref{sec:parameter_reduction}, where we describe the method in general, in particular to \autoref{sec:EfficientSU2_again} for specific details of the analysis regarding \verb|EfficientSU2(3, reps=N)|. For a hands-on example that incorporates additional translational symmetry, we refer to \autoref{sec:4Q_circuit_design}. The technical details of incorporating such symmetries are given in \autoref{sec:translation_invariance}. 

\subsection{Removing interdependent parameters}

The state space of a three-qubit quantum device has real dimension~$15$, because it is the unit sphere of a $2^3$-dimensional complex Hilbert space. Hence, a circuit without interdependent parameters has no more than $15$ parameters, i.e., \verb|EfficientSU2(3, reps=N)| with $N\ge2$ must have at least three parameters that do not contribute to the circuits expressivity. Of course, in the $N=2$ case, it is not possible to remove any three gates arbitrarily. For example, removing the three initial $R_Y$ rotations renders the entire first layer ineffective. Thus, it effectively removes six gates and the resulting circuit cannot be maximally expressive. Similarly, there is no unique choice of parameter removal, and different applications might require different choices to be made. Dimensional expressivity analysis therefore aims to identify the parameters that need to be removed. In particular, gate prioritization will be discussed in \autoref{sec:gate_priority}. A more reasonable reduction to $15$ parameters would be to remove the last three $R_Z$ gates and consider the circuit given in the lower panel of \autoref{fig:EffSU2}. For most parameter sets, this will indeed provide a maximally expressive circuit. \verb|EfficientSU2(3, reps=2)| therefore provides an interesting test bed of the dimensional expressivity analysis, because it is simple enough to verify the analysis results, while being sufficiently complex to obtain interesting insights.

Dimensional expressivity analysis (\autoref{sec:EfficientSU2_again}) using random parameter values $\theta_1,\ldots,\theta_{18}$ shows that \verb|EfficientSU2(3, reps=2)| is maximally expressive with $15$ independent parameters (that is, the circuit is a locally surjective map into the state space). In fact, the circuit can be reduced to the circuit given in the lower panel of \autoref{fig:EffSU2} which is still maximally expressive and only contains independent parameters. In this sense, \verb|EfficientSU2(3, reps=2)| with global choice $\theta_{16}=\theta_{17}=\theta_{18}=0$ is the most efficient, maximally expressive circuit in the \verb|EfficientSU2(3, reps=N)|-family. In particular, it is not advisable to use \verb|EfficientSU2(3, reps=N)| with $N>2$ as it introduces additional parameters (and thus additional work for the optimizer) without increasing the expressivity of the circuit.

However, this maximal expressivity property is not true for all parameter sets (cf. \autoref{sec:semicontinuity}). Some singular parameter sets exist. For example, a common starting position for a variational quantum simulation is likely to be $\theta_{\rm initial}=0$, i.e., all $\theta_j$ are initially set to $0$. But starting with $\theta_{\rm initial}=0$, there are only seven independent parameters $\theta_1$, $\theta_2$, $\theta_3$, $\theta_4$, $\theta_7$, $\theta_8$, and $\theta_{13}$. This reduction is most easily seen with $\theta_4$, $\theta_5$, and $\theta_6$. If $\theta_1=\theta_2=\theta_3=0$, then $R_Z(\theta_4)$, $R_Z(\theta_5)$, and $R_Z(\theta_6)$ all act on their respective qubit which is in the state $\ket0$. In other words, they all have the same effect of introducing a relative phase. As such, the three parameters are mutually dependent and we consider $\theta_4$ to be the ``independent'' parameter because it is the first parameter (with respect to the chosen ordering of parameters) that generates arbitrary phases.

This effect can be detrimental to local optimizers. In particular, gradient based optimizers using \verb|EfficientSU2(3, reps=2)| starting with $\theta=0$ might be trapped in an artificial local minimum purely because a variation of the seven independent parameters remains inside a level set\footnote{A level set $L_{c}(f)$ of a function $f$ of $n$ variables is a set where the function takes on a given constant value $c$: $L_{c}(f)=\left\{(x_{1},\cdots ,x_{n})\,\mid \,f(x_{1},\cdots ,x_{n})=c\right\}$. } of the energy landscape and hence the gradient is rendered zero. Replacing the starting value $\theta_{\rm initial}=0$ with a small random perturbation $\theta_{\rm initial}=\eps$ is unlikely to result in such an artificial minimum and provides the full dimensional expressivity of the circuit.

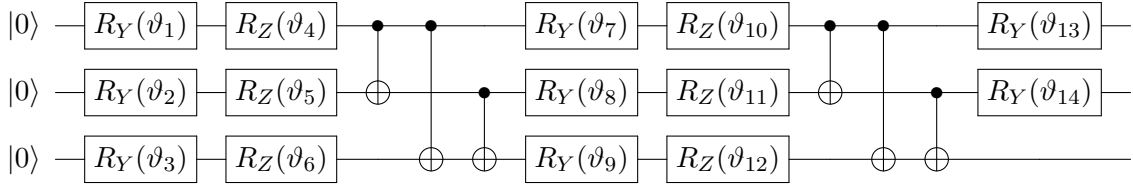
\begin{figure*}[!ht]
  \begin{align*}
      \Qcircuit @C=1em @R=.7em {
        \lstick{\ket{0}} & \gate{R_Y(\theta_1)} & \gate{R_Z(\theta_4)} & \ctrl{1} & \ctrl{2} & \qw      & \gate{R_Y(\theta_7)} & \gate{R_Z(\theta_{10})} & \ctrl{1} & \ctrl{2} & \qw      & \gate{R_Y(\theta_{13})} & \qw \\
        \lstick{\ket{0}} & \gate{R_Y(\theta_2)} & \gate{R_Z(\theta_5)} & \targ    & \qw      & \ctrl{1} & \gate{R_Y(\theta_8)} & \gate{R_Z(\theta_{11})} & \targ    & \qw      & \ctrl{1} & \gate{R_Y(\theta_{14})} & \qw\\
        \lstick{\ket{0}} & \gate{R_Y(\theta_3)} & \gate{R_Z(\theta_6)} & \qw      & \targ    & \targ    & \gate{R_Y(\theta_9)} & \gate{R_Z(\theta_{12})} & \qw      & \targ    & \targ    & \qw & \qw
        }
  \end{align*}
  \caption{Reduction of QISKIT's \texttt{EfficientSU2} 2-local circuit similar to the lower panel of \autoref{fig:EffSU2}, but with removing the option to change a global phase.}\label{fig:Eff.SU2-red2}
\end{figure*} 

In the light of these singular sets, for which the dimensional expressivity is not maximal, it is interesting to note that performing the dimensional expressivity analysis on random parameters suffices to make general statements. This is a consequence of the generic structure of this type of circuit. The singular parameters are precisely the zeros of a function comprised of sums and products of sines and cosines (cf. \autoref{sec:semicontinuity}). In other words, they are a lower dimensional subset of the parameter space. This implies that drawing\footnote{with respect to some measure that is absolutely continuous with respect to the volume measure on the parameter space (Lebesgue or Haar measure in most cases)} random parameters avoids this set of singular parameters with probability $1$. Fundamentally, this is the reason why a random perturbation on the initial parameter set (like $\theta_{\rm initial}=0$ here) avoids artificial minima with probability $1$ as well.

\subsection{Removing global phases}

The $\theta_{\rm initial}=0$ example also highlights a particular curiosity of the dimensional expressivity analysis. Above, we considered $\theta_4$ to be an independent and ``relevant'' parameter because it introduces a global phase. Mathematically, $\theta_4$ is necessary because it is impossible to generate an arbitrary vector in the unit sphere of the $2^3$-dimensional complex Hilbert space if it is not possible to generate arbitrary phases. However, for quantum computing applications, such a global phase is irrelevant. We would therefore be interested in removing any gates or parameters whose only effect is a global phase change.

Assuming we wish to analyze a circuit~$C(\theta)$ 
\begin{align*}
  \Qcircuit @C=1em @R=.7em {
    \lstick{\ket{0}} & \multigate{3}{C(\theta)} &\qw\\
    \lstick{\ket{0}} & \ghost{C(\theta)} &\qw\\
    \vdots & \nghost{C(\theta)} &\vdots\\
    \lstick{\ket{0}} & \ghost{C(\theta)} &\qw\\
  }
\end{align*}
we may instead analyze the circuit $\tilde C(\phi,\theta)$, which is given by $C(\theta)$ with an additional $R_Z(\phi)$ gate:
\begin{align*}
  \Qcircuit @C=1em @R=.7em {
    \lstick{\ket{0}} & \gate{R_Z(\phi)} & \multigate{3}{C(\theta)} &\qw\\
    \lstick{\ket{0}} & \qw & \ghost{C(\theta)} &\qw\\
    & \vdots & \nghost{C(\theta)} &\vdots\\
    \lstick{\ket{0}} & \qw & \ghost{C(\theta)} &\qw\\
  }
\end{align*}
The additional gate in $\tilde C$ forces the parameter~$\phi$ to be independent (being that it is the first parameter) and its effect is the introduction of a relative phase. Due to compactness of the parameter space and lower semi-continuity of parameter independence (cf.,~\autoref{sec:semicontinuity}), checking this relative phase generation is sufficient to conclude about global phase symmetry of the circuit (cf.,~\autoref{sec:symmetry_removal}). Therefore, in the remaining part of the paper, we will call such phases ``global'' independently of whether they are applied to a single qubit or multiple qubits.

When analyzing $\tilde C$ instead of $C$, we may encounter a situation in which adding $\theta_j$ to the circuit $C$ enlargens the set of reachable states (which is why it is considered an independent parameter) but the use of $\theta_j$ contributes at most a global phase factor to the set of reachable states (which is why we would want to consider $\theta_j$ as a dependent parameter). More precisely, dimensional expressivity analysis of $C$ yields independence of the parameter $\theta_j$ but every state $\ket\psi$, that can be reached with $C$ using only the parameters $\theta_1$, $\ldots$, $\theta_j$, is of the form $e^{i\alpha}\ket{\psi'}$ where $\ket{\psi'}$ can be reached with $C$ using only the parameters $\theta_1$, $\ldots$, $\theta_{j-1}$. This precise situation presents itself when analyzing $\theta_4$ in \verb|EfficientSU2(3, reps=2)| at $\theta_{\rm initial}=0$. Analyzing $\tilde C$ instead of $C$, such global phases can already be generated using the parameters $\phi$, $\theta_1$, $\ldots$, $\theta_{j-1}$. The addition of $\theta_j$ therefore no longer enlargens the set of reachable states and $\theta_j$ will be identified as a dependent parameter.

\begin{figure*}[!ht]
  \begin{align*}
      \Qcircuit @C=1em @R=.7em {
        \lstick{\ket{0}} & \gate{R_Y(\theta_1)} & \gate{R_Z(\theta_4)} & \ctrl{1} & \ctrl{2} & \qw      & \gate{R_Y(\theta_7)} & \ctrl{1} & \ctrl{2} & \qw      & \gate{R_Y(\theta_{13})} & \gate{R_Z(\theta_{16})} & \qw \\
        \lstick{\ket{0}} & \gate{R_Y(\theta_2)} & \gate{R_Z(\theta_5)} & \targ    & \qw      & \ctrl{1} & \gate{R_Y(\theta_8)} & \targ    & \qw      & \ctrl{1} & \gate{R_Y(\theta_{14})} & \gate{R_Z(\theta_{17})} & \qw\\
        \lstick{\ket{0}} & \gate{R_Y(\theta_3)} & \gate{R_Z(\theta_6)} & \qw      & \targ    & \targ    & \qw & \qw      & \targ    & \targ    & \gate{R_Y(\theta_{15})} & \gate{R_Z(\theta_{18})} & \qw
      }
  \end{align*}
  \caption{Reduction of QISKIT's \texttt{EfficientSU2} 2-local circuit similar to \autoref{fig:Eff.SU2-red2}, but with different gate ordering.}\label{fig:Eff.SU2-red3}
\end{figure*}
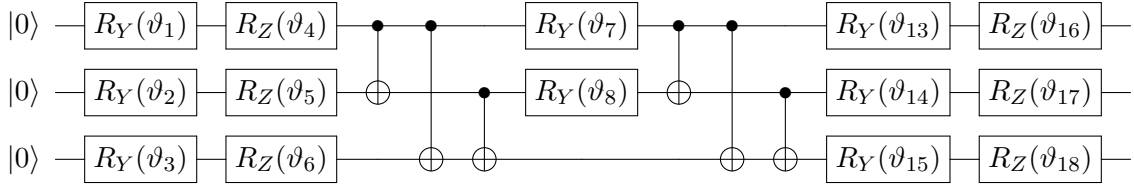 

In application to \verb|EfficientSU2(3, reps=2)| at $\theta_{\rm initial}=0$, we therefore obtain the six independent parameters $\theta_1$, $\theta_2$, $\theta_3$, $\theta_7$, $\theta_8$, and $\theta_{13}$. Application to \verb|EfficientSU2(3, reps=2)| at random values of $\theta$ shows that for all tested values the parameter $\theta_{15}$ adds the global phase. Thus, if we globally set $\theta_{15}=0$ as well, and only consider the circuit in \autoref{fig:Eff.SU2-red2}, we still have a maximally expressive circuit -- in the sense relevant to quantum computing -- and we have optimized the circuit by removing the option to only change a global phase factor.

Of course, as an analysis tool, this idea can be generalized to other symmetries we wish to avoid. As long as we can add a one parameter block $U$
\begin{align*}
  \Qcircuit @C=1em @R=.7em {
    \lstick{\ket{0}} & \multigate{2}{U(\phi)} & \multigate{2}{C(\theta)} &\qw\\
    \lstick{\vdots} & \nghost{U(\phi)} & \nghost{C(\theta)} &\vdots\\
    \lstick{\ket{0}} & \ghost{U(\phi)} & \ghost{C(\theta)} &\qw\\
  }
\end{align*}
we can remove parameters from the quantum circuit which only add a symmetry that is to be avoided. We will discuss this procedure in more detail in \autoref{sec:symmetry_removal}.

\subsection{Gate and parameter priority in dimensional expressivity analysis}\label{sec:gate_priority}
The final circuit given in the lower panel of \autoref{fig:EffSU2} is maximally expressive and has no redundant parameters. It is therefore optimal from the dimensional expressivity analysis point of view. However, it may not be the optimal reduced circuit for certain VQS applications. For example, it may be useful to have a circuit that can be easily initialized in an arbitrary product state. A possibility for such a circuit is the original \verb|EfficientSU2(3, reps=N)|. Setting all parameters except the final layer to zero, reduces the circuit to
\begin{align*}
  \Qcircuit @C=1em @R=.7em {
    \lstick{\ket{0}} & \multigate{2}{\id} & \gate{R_Y(\theta_{6N+1})} & \gate{R_Z(\theta_{6N+4})} & \qw \\
    \lstick{\ket{0}} & \ghost{\id} & \gate{R_Y(\theta_{6N+2})} & \gate{R_Z(\theta_{6N+5})} & \qw\\
    \lstick{\ket{0}} & \ghost{\id} & \gate{R_Y(\theta_{6N+3})} & \gate{R_Z(\theta_{6N+6})} & \qw
  }
\end{align*}
which is maximally expressive and free from redundant parameters for product states (this will be shown in \autoref{sec:circuit_reduction_single_qubit}). 

However, generating arbitrary product states is at first sight not as easy for the circuit in \autoref{fig:Eff.SU2-red2} as for \verb|EfficientSU2(3, reps=N)|.  If we want to ensure that the reduced circuit can generate product states just as easily as \verb|EfficientSU2(3, reps=N)|, then we need to ensure the same $R_Y R_Z$ structure at the end of the reduced circuit. Luckily, we can adjust the dimensional expressivity analysis slightly in order to force this structure of the reduced circuit.

If $\theta_j$ and $\theta_k$ are mutually dependent and $j<k$, then the dimensional expressivity analysis will consider $\theta_j$ to be ``independent'' and $\theta_k$ ``redundant''. This is based on the implicit parameter priority order $j<k$, i.e, ``$j$ is more important than $k$''. In general, let us consider a circuit with the structure
\begin{align*}
  \Qcircuit @C=1em @R=.7em {
    &\multigate{2}{U_1} & \multigate{2}{G(\theta_j)} & \multigate{2}{U_2} & \multigate{2}{G'(\theta_k)} & \multigate{2}{U_3} &\qw\\
    \lstick{\vdots} &\nghost{U_1} & \nghost{G(\theta_j)} & \nghost{U_2} & \nghost{G'(\theta_k)} & \nghost{U_3} &\vdots\\
    &\ghost{U_1} & \ghost{G(\theta_j)} & \ghost{U_2} & \ghost{G'(\theta_k)} & \ghost{U_3} &\qw\\
  }
\end{align*}
with mutually dependent $\theta_j$ and $\theta_k$ and $j<k$. If we wish to prioritize the gate block $G'$ over the gate block $G$, then we can swap $\theta_j$ and $\theta_k$ to analyze the circuit
\begin{align*}
  \Qcircuit @C=1em @R=.7em {
    &\multigate{2}{U_1} & \multigate{2}{G(\theta_k)} & \multigate{2}{U_2} & \multigate{2}{G'(\theta_j)} & \multigate{2}{U_3} &\qw\\
    \lstick{\vdots} &\nghost{U_1} & \nghost{G(\theta_k)} & \nghost{U_2} & \nghost{G'(\theta_j)} & \nghost{U_3} &\vdots\\
    &\ghost{U_1} & \ghost{G(\theta_k)} & \ghost{U_2} & \ghost{G'(\theta_j)} & \ghost{U_3} &\qw\\
  }
\end{align*}
This change only affects the implicit gate priority as introduced by the ordering of parameters. The circuit itself is unchanged (provided the ranges for $\theta_j$ and $\theta_k$ are swapped as well). Dimensional expressivity analysis will still consider $\theta_j$ to be ``independent'' but now this means keeping $G'$ rather than $G$.

Assuming we wish to prioritize the last layer in \verb|EfficientSU2(3, reps=N)|, we can relabel the parameters to give the structure
\begin{align*}
  \Qcircuit @C=1em @R=.7em {
    \lstick{\ket{0}} & \multigate{2}{U(\hat\theta)} & \gate{R_Y(\theta_{1})} & \gate{R_Z(\theta_{4})} & \qw \\
    \lstick{\ket{0}} & \ghost{U(\hat\theta)} & \gate{R_Y(\theta_{2})} & \gate{R_Z(\theta_{5})} & \qw\\
    \lstick{\ket{0}} & \ghost{U(\hat\theta)} & \gate{R_Y(\theta_{3})} & \gate{R_Z(\theta_{6})} & \qw
  }
\end{align*}
where $\hat\theta:=(\theta_7,\ldots,\theta_{6N+6})$. Repeating the dimensional expressivity analysis (including the removal of a global phase) with this relabeling of the parameters (and returning to the initial parameter labeling) yields the maximally expressive redundancy free circuit in \autoref{fig:Eff.SU2-red3}. Both circuits in \autoref{fig:Eff.SU2-red2} and \autoref{fig:Eff.SU2-red3} are maximally expressive and free from redundant parameters. From the dimensional expressivity analysis point of view they are therefore equivalent. For any given application this choice of gate and parameter prioritization may matter.

\section{Circuit parameter reduction and manifold dimension}\label{sec:parameter_reduction}
In order to quantify the expressivity of a given quantum circuit, we consider the dimension of the manifold of states that can be reached by the circuit. This choice is based on the fact that many quantum simulations use optimization loops to find optimal parameter sets for the quantum circuit. While it is possible for a low dimensional submanifold of reachable states to approximate every point of the state space (e.g., a dense spiral on the Bloch sphere), having a large codimension increases the likelihood of two states close in state space to be very far apart in parameter space. Hence, high codimension increases the likelihood of a local optimizer in a variational quantum simulation to fail.

Unfortunately, many quantum circuits have a large number of interdependent parameters. This means that each state can be reached through numerous choices of parameter values. In order to compute the dimension of the manifold of states reachable by a given quantum circuit, it is necessary to restrict the parameter space in such a way that the dimension can be extracted. This furthermore has the advantage that we can identify a minimal set of independent parameters. Optimizing only the independent parameters thus reduces computational effort and knowledge of the independent parameters can be used as a guideline in circuit design. Similarly, knowing the codimension of the manifold of reachable states leads us to a natural measure of deficiency for a given quantum circuit. In circuit design, minimizing this deficiency will lead to maximal expressivity at a minimal number of parameters to optimize.

In \autoref{sec:circuit_reduction_general_method} we will describe this parameter reduction in general terms. Of course, up to this point we have been assuming that the states reachable by a given quantum circuit form a manifold. This is not obvious a~priori but we will show that this will at least be locally true, provided the conditions assumed in \autoref{sec:circuit_reduction_general_method} are satisfied. In \autoref{sec:semicontinuity}, we will discuss the local behavior of the parameter reduction and obtain estimates on the validity of reduced circuits, that is, how much variation of parameters is possible for any given reduction of the parameters. We will then consider a single-qubit example in \autoref{sec:circuit_reduction_single_qubit} for explicit illustration and return to QISKIT's EfficientSU2 2-local circuit \verb|EfficientSU2(3, reps=N)| in \autoref{sec:EfficientSU2_again} to provide the details of the analysis in \autoref{sec:EfficientSU2}. Finally, in \autoref{sec:symmetry_removal} we will discuss the removal of symmetries procedure that already allowed us to reduce \verb|EfficientSU2(3, reps=N)| to a circuit that can generate arbitrary states up to a global phase. 

\subsection{General method}\label{sec:circuit_reduction_general_method}
For the remainder of this paper, it is advantageous to consider a quantum circuit to be a map $C:\ P\to\d B_{\Hp}$ where $P$ is the parameter space and $\d B_{\Hp}$ the unit sphere in the Hilbert space $\Hp$. For a $Q$-qubit system, $\Hp=\cn^{2^Q}$. It is important to note that this definition of a quantum circuit now includes the evaluation on the initial state. For example, if the initial state of a single-qubit system is $\ket0$ and the gate sequence is $R_X(\theta)$, then
\begin{align}
    C(\theta)=&R_X(\theta)\ket0=\cos\frac\theta2\ket0-i\sin\frac\theta2\ket1.
\end{align}
If $P$ is a manifold (e.g., if the circuit contains $N$ rotation gates and the entangling layers do not contain any parameters, $P$ will be a flat torus $(\rn/2\pi\zn)^N$) and $P$ locally contains a submanifold $P_C$ such that $\check C:=C|_{P_C}:\ P_C\to\d B_{\Hp}$ is a diffeomorphism, then $C$ defines a submanifold of $\d B_{\Hp}$. To show that this is indeed the case, the important questions to ask are
\begin{enumerate}
\item[(i)] Does $C$ locally define a submanifold of $\d B_{\Hp}$?
\item[(ii)] If $C$ locally defines a submanifold, what is its codimension?
\item[(iii)] If $C$ locally defines a submanifold, can we restrict the parameter space to a minimal set?
\end{enumerate}
Although the image of $C$ is naturally described as a subset of a complex Hilbert space, because the parameter space is real, we have to treat the image of $C$ as a real manifold. Furthermore, the questions (i), (ii), and (iii) are not independent of each other. In fact,  the best way to show that we have a manifold already requires the restricted circuit $\check C$. Hence, it is advantageous to start with the construction of the parameter space restriction assuming that the image of $C$ is a manifold and then show that the image of the restricted parameter set is already locally surjective on the image of $C$.

To restrict the parameter space locally, we can look at the tangent space of the image. If we have $N$ parameters, then the tangent space is spanned by $\d_1C,\ldots,\d_NC$. Thus, it is interesting to consider the Jacobian
\begin{align}
  C'(\theta)=
  \begin{pmatrix}
    |&|&&|\\
    \d_1C(\theta)&\d_2C(\theta)&\ldots&\d_NC(\theta)\\
    |&|&&|
  \end{pmatrix}.
\end{align}
The rank of $C'(\theta)$ therefore gives us the number of independent parameters of $C$ at $\theta\in P$. Furthermore, if we consider $J_\theta\sse\{1,\ldots,N\}$ such that $\{\d_jC(\theta);\ j\in J_\theta\}$ is a maximal linearly independent subset of $\{\d_jC(\theta);\ 1\le j\le N\}$, then we can define the restriction $\check C_\theta$ in a neighborhood of $\theta$ by keeping $\theta_j$ constant whenever $j\notin J_\theta$. This implies that the columns of the Jacobian $\check C_\theta'(\theta)$ are precisely the $\d_jC(\theta)$ with $j\in J_\theta$. In other words, $\check C_\theta$ is locally an immersion and any state reached by variation of parameters $\theta_j$ with $j\notin J_\theta$ can also be reached varying the $\theta_k$ with $k\in J_\theta$. To an abstract degree, this answers all three questions. 
\begin{enumerate}
\item[(i)] The restricted circuit $\check C_\theta$ locally defines a submanifold of $\d B_{\Hp}$ since it is an immersion and an immersion is locally an embedding. 
\item[(ii)] The dimension of the manifold locally defined through $\check C_\theta$ has the dimension $\#J_\theta$. Thus, the codimension of the circuit $\check C_\theta$ is $\dim\Hp-1-\#J_\theta$ (where $\Hp$ is the $2^Q$-dimensional Hilbert space of a $Q$-qubit system).
\item[(iii)] The restriction to $\check C_\theta$ is minimal, that is, any further restriction of the circuit or parameter space will not be surjective on the image of the original circuit anymore and, therefore, reduce expressivity of the circuit.
\end{enumerate}

\subsection{Lower-semicontinuity and constancy of independent parameters}\label{sec:semicontinuity}

In this section, we will discuss the local behavior of the parameter reduction procedure. This will be of fundamental importance for estimating the set of parameters on which any given restricted quantum circuit is valid and for identification of singular points at which the dimensional expressivity of a given quantum circuit is reduced.

It is important to note that the circuit manifold dimension is lower semi-continuous, that is, if the restricted circuit $\check C_\theta$ defines a manifold of dimension $n$ in a neighborhood of $\theta$ then the image of $C$ has dimension \emph{at least} $n$. It is possible that a small perturbation of $\theta$ increases expressivity but it cannot decrease.

This claim follows from lower-semicontinuity of $\rank C'(\theta)$ because the dimension of the manifold defined by $\check C_\theta$ is precisely the rank of $C'(\theta)$. This can be proven directly by assuming $\rank C'(\theta)=:r$. Then there exists a non-singular $r\times r$-submatrix $C_0$ of $C'(\theta)$. Thus $\det C_0\ne0$ and continunity of the determinant implies that for any sufficiently small perturbation $\tilde C$ of $C'(\theta)$ the $r\times r$-submatrix of $\tilde C$ corresponding to the same indices as $C_0$ must be non-singular as well. In other words, $\rank\tilde C\ge \rank C'(\theta)$ holds and lower-semicontinuity is proven.

Conversely, it is possible for the dimension to jump upwards under small perturbations. This behavior has already been observed in \autoref{sec:EfficientSU2}. An explicit example can be found in \verb|EfficientSU2(2, reps=0)|, i.e., the circuit 
\begin{align}
  \begin{split}
    C(\theta):=&\, R_{Z_2}(\theta_4)R_{Z_1}(\theta_3)R_{Y_2}(\theta_2)R_{Y_1}(\theta_1)\ket{00}\\
    =&\, e^{-i\frac{\theta_3+\theta_4}{2}}\cos\frac{\theta_2}{2}\cos\frac{\theta_1}{2}\ket{00}\\
    &+e^{-i\frac{-\theta_3+\theta_4}{2}}\cos\frac{\theta_2}{2}\sin\frac{\theta_1}{2}\ket{01}\\
    &+e^{-i\frac{\theta_3-\theta_4}{2}}\sin\frac{\theta_2}{2}\cos\frac{\theta_1}{2}\ket{10}\\
    &+e^{i\frac{\theta_3+\theta_4}{2}}\sin\frac{\theta_2}{2}\sin\frac{\theta_1}{2}\ket{11}.
  \end{split}
\end{align}
If $\theta_1=\theta_2=0$, then $\d_3C(0,0,\theta_3,\theta_4)$ and $\d_4C(0,0,\theta_3,\theta_4)$ coincide, reducing the manifold dimension of $\check C_{(0,0,\theta_3,\theta_4)}$ to $3$. However, this is not the case for any small perturbation in $(\theta_1,\theta_2)$.

This lower semi-continuity property of the circuit reduction indicates that it may be advantageous to introduce small random perturbations if a quantum simulation appears stuck in what seems to be a local minimum. In particular, if an initial parameter set is generated by some analytic method, an implementation might be improved through a small random perturbation of the initial parameter set.

The lower-semicontinuity property has shown that a small perturbation of the parameters can increase the number of independent parameters and that independent parameters stay independent under small perturbation. Crucially, it also allows us to estimate how large such a perturbation may be. Let   $C_0(\theta_0)$ be the maximal invertible submatrix as determined in the lower-semicontinuity proof above.  We will denote the $\rank(C_0(\theta_0))\times\rank(C_0(\theta_0))$-submatrix of $C'(\theta)$ corresponding to the same indices as $C_0(\theta)$ for any parameter $\theta$.\footnote{To illustrate this submatrix construction, consider $C'(\theta):=\begin{pmatrix}a(\theta)&b(\theta)&c(\theta)\\d(\theta)&e(\theta)&f(\theta)\\g(\theta)&h(\theta)&i(\theta)\end{pmatrix}$. At some value $\theta_0$, point evaluation may yield $C'(\theta_0)=\begin{pmatrix}1&1&1\\1&1&1\\1&2&1\end{pmatrix}$. This matrix is of rank $2$ and has the invertible $2\times2$-submatrix $C_0(\theta_0)=\begin{pmatrix}1&1\\1&2\end{pmatrix}=\begin{pmatrix}a(\theta_0)&b(\theta_0)\\g(\theta_0)&h(\theta_0)\end{pmatrix}$. The submatrix of $C'(\theta)$ corresponding to the same indices as $C_0(\theta_0)$ is then given by $C_0(\theta)=\begin{pmatrix}a(\theta)&b(\theta)\\g(\theta)&h(\theta)\end{pmatrix}$.}

Continuity of $\det\circ C_0$ implies the existence of $R>0$ such that 
\begin{align}
  \fa \theta\in B(\theta_0,R):\ \det(C_0(\theta))\ne0
\end{align}
holds. This implies that the independent parameters identified for $\theta_0$ remain independent for all $\theta$ in the open ball $B(\theta_0,R):=\{\theta;\ \dist(\theta,\theta_0)<R\}$ where $\dist$ denotes the distance in parameter space. We can estimate such a value of $R$ by choosing an initial $R_0>0$ and computing the derivative of $\det\circ C_0$ on the closed ball $B[\theta_0,R_0]$. Setting $\delta:=\max_{\theta\in B[\theta_0,R_0]}\norm{(\det\circ C_0)'(\theta)}_{\ell_2}$ where $\norm{(\det\circ C_0)'(\theta)}_{\ell_2}$ denotes the $\ell_2$-norm of the gradient $(\det\circ C_0)'(\theta)$, we obtain the estimate $R\ge\min\{R_0,\delta^{-1}\abs{\det(C_0(\theta_0))}\}$.

We will call this $R$ the radius of validity for a reduced circuit. It is of particular importance because it guarantees that no singular points are closer than $R$ to the parameter set at which a reduced circuit was constructed. As such, can be used to anticipate problems arising from a loss of dimensional expressivity. We will therefore make heavy use of the radius of validity in \autoref{sec:circuit_reduction_single_qubit}, where we will construct reduced circuits for a single-qubit example, and in \aref{app:eigenspaces_path_connected}, where it will be used to identify cases in which a quantum simulation can fail due to loss of dimensional expressivity. \autoref{fig:overlapping_radii_of_validity_regular} and \autoref{fig:overlapping_radii_of_validity_singular} in particular show how the radii of validity behave on paths with and without singular points.

To illustrate this estimation of the radius of validity, let us consider the single-qubit circuit 
\begin{align}
  C(\theta):=R_X(\theta)\ket0=\cos\frac{\theta}{2}\ket0-i\sin\frac{\theta}{2}\ket1.
\end{align}
The Jacobian is therefore given by
\begin{align}
  C'(\theta)= \frac{1}{2}
  \begin{pmatrix}
    -\sin\frac\theta2\\
    -i\cos\frac\theta2
  \end{pmatrix}
  .
\end{align}
However, this Jacobian is over $\cn$ whereas our analysis is over $\rn$. We therefore need to split real and imaginary parts, doubling the dimension, and yielding
\begin{align}
  C'(\theta)= \frac{1}{2}
  \begin{pmatrix}
    -\sin\frac\theta2\\
    0\\
    0\\
    -\cos\frac\theta2
  \end{pmatrix}
  .
\end{align}
For any point with $-\frac12\sin\frac\theta2\ne0$, the submatrix $C_0$ can be chosen as the $1\times1$-matrix $\l(-\frac{1}{2}\sin\frac\theta2\r)$. Hence, $\det(C_0(\theta))=-\frac{1}{2}\sin\frac\theta2$ and $(\det\circ C_0)'(\theta)=-\frac{1}{4}\cos\frac\theta2$ imply $\delta\le\frac14$ for any choice of $R_0$ as well as the estimate $R\ge2\abs{\sin\frac\theta2}$ provided $R_0\ge2$ is chosen.

Unfortunately, the estimate for the radius of validity $R$ for the reduction to independent parameters converges to $0$ as $\frac\theta2$ converges to a zero of the sine. This limitation is artificial since the one parameter in the circuit is always independent which we can also see switching to $-\frac{1}{2}\cos\frac\theta2$ near zeros of the sine. This problem can be addressed using the fact that $\rank(A)=\rank(A^*A)$ holds for all matrices $A$. Thus, if we choose $\tilde C_0$ to be the matrix of all independent columns of $C'(\theta)$, i.e.,
\begin{align}
    \tilde C_0=& \frac{1}{2}
    \begin{pmatrix}
      -\sin\frac\theta2\\
      0\\
      0\\
      -\cos\frac\theta2
    \end{pmatrix}
    ,
\end{align}
then we obtain $\det(\tilde C_0^*\tilde C_0)=\frac14$ for all $\theta$. This shows directly that the circuit has only independent parameters and that this is valid for all parameters.

In practical terms, the lower-semicontinuity property and the corresponding radius of validity imply that the pointwise performed dimensional expressivity analysis is valid locally. More precisely, if we choose a point $\Theta$ in parameter space, identify the set independent parameters $\Theta_{i_1},\ldots,\Theta_{i_k}$ at that point, and compute the radius of validity $R$ at $\Theta$, then all points $\theta$ with $\dist(\theta,\Theta)<R$ have independent parameters $\theta_{i_1},\ldots,\theta_{i-k}$. In other words, independent parameters remain independent under small perturbations on point at which the dimensional expressivity analysis is performed. 

Let us furthermore assume that the parameter space $P$ is a compact manifold, as is the case for parametric quantum circuits containing only rotation gates and non-parametric gates. Then, we can use this perturbation resistance of independent parameters to find a finite set of points that need to be checked such that a set of independent parameters is known for every point in parameter space. To formalize this, let us define $I_\Theta$ as the set of independent parameters obtained from dimensional expressivity analysis at a point $\Theta\in P$, i.e., the parameters $\Theta_i$ with $i\in I_\Theta$ are independent. Let $R_\Theta$ be the corresponding radius of validity. Then, by lower-semicontinuity, the parameters $\theta_i$ with $i\in I_\Theta$ are independent whenever $\dist(\theta,\Theta)<R_\Theta$. However, the open balls $B(\Theta,R_\Theta)=\{\theta\in P;\ \dist(\theta,\Theta)<R_\Theta\}$ form an open cover of $P$. By compactness, there exists a finite subcover, i.e., there exist finitely many $\Theta^{(1)},\ldots,\Theta^{(n)}$ such that $P=\bigcup_{j=1}^nB(\Theta^{(j)},R_{\Theta^{(j)}})$. Given any $\theta\in P$, we can therefore find a $j_\theta$ such that $\theta\in B(\Theta^{(j_\theta)},R_{\Theta^{(j_\theta)}})$ and we conclude that the parameters $\theta_i$ with $i\in I_{\Theta^{(j_\theta)}}$ are independent. Hence, in order to find sets of independent parameters at every point in parameter space, it suffices to perform the dimensional expressivity analysis at finitely many points.

While this allows for the construction of finitely many minimal and maximally expressive quantum circuits that cover the entire parameter space, it should be noted that it may not always be prudent to construct a finite family of circuits, but to allow for an infinite family of circuits to exist on this finite cover. This leads to the notion of ``seamless switching'' as discussed in \autoref{sec:finitely_many_circuits} and is, more generally, related to path connectivity questions (cf.~\aref{app:eigenspaces_path_connected}). It should also be noted that testing sufficiently many points in parameter space to obtain an a-priori atlas of independent parameters may not always be necessary. In some applications, it may be sufficient to choose a general method of constructing the circuit and performing the dimensional expressivity analysis for each update of a VQS. In other words, the idea would be to construct a set of independent parameters for the initial set of VQS parameters and to use the corresponding restricted circuit until the parameter update maps outside the current radius of validity. Then, a new circuit can be constructed and used until the VQS leaves its radius of validity. Of course, such an on the fly analysis requires an efficient and automated implementation which will be discussed in \autoref{sec:implementation}. 

\subsection{Single-qubit case example}\label{sec:circuit_reduction_single_qubit}
In this section, we will illustrate the dimensional expressivity analysis on a single-qubit example. We will start our discussion of the single-qubit case analyzing a simple two-parameter circuit in \autoref{sec:2_param_circuit}. Since this first two-parameter circuit cannot be maximally expressive (that requires three parameters), we will then consider an artificial four-parameter circuit in \autoref{sec:4_param_circuit} to highlight how to identify interdependent parameters and explicitly reduce the circuit. This will yield multiple reduced circuits highlighting a behavior that is generically to be expected. In \autoref{sec:finitely_many_circuits}, we therefore discuss this generic behavior giving us multiple locally reduced circuits in more detail. We also prove that finitely many reduced circuits are sufficient to cover the entire parameter space provided the parameter space is compact. Finally, in \autoref{sec:4_param_circuit_global_phase} we will return the reduced four-parameter circuits of \autoref{sec:4_param_circuit} and apply some symmetry removal techniques. These will be developed in \autoref{sec:symmetry_removal} but at this point, they will already allow us to further simplify the reduced circuits of \autoref{sec:4_param_circuit}.

\subsubsection{A two-parameter circuit}\label{sec:2_param_circuit}
The first circuit we may consider is
\begin{align}
  C(\theta):=R_Z(\theta_2)R_X(\theta_1)\ket0
\end{align}
with parameter space $(\rn/2\pi\zn)^2$. In the coordinates of the Hilbert space $\cn^2$, $C(\theta)$ can thus be represented as
\begin{align}
  C(\theta)=&
  \begin{pmatrix}
    \cos\frac{\theta_1}{2}\cos\frac{\theta_2}{2}-i\cos\frac{\theta_1}{2}\sin\frac{\theta_2}{2}\\
    -i\sin\frac{\theta_1}{2}\cos\frac{\theta_2}{2}+\sin\frac{\theta_1}{2}\sin\frac{\theta_2}{2}
  \end{pmatrix}
  .
\end{align}
This implies
\begin{align}
  \d_1C(\theta)=&\frac{1}{2}
  \begin{pmatrix}
    -\sin\frac{\theta_1}{2}\cos\frac{\theta_2}{2}+i\sin\frac{\theta_1}{2}\sin\frac{\theta_2}{2}\\
    -i\cos\frac{\theta_1}{2}\cos\frac{\theta_2}{2}+\cos\frac{\theta_1}{2}\sin\frac{\theta_2}{2}
  \end{pmatrix}
\end{align}
and
\begin{align}
  \d_2C(\theta)=&\frac{1}{2}
  \begin{pmatrix}
    -\cos\frac{\theta_1}{2}\sin\frac{\theta_2}{2}-i\cos\frac{\theta_1}{2}\cos\frac{\theta_2}{2}\\
    i\sin\frac{\theta_1}{2}\sin\frac{\theta_2}{2}+\sin\frac{\theta_1}{2}\cos\frac{\theta_2}{2}
  \end{pmatrix}
  .
\end{align}
Since $C$ is not constant, its Jacobian is either of rank $1$ or $2$. For the Jacobian to be of rank $1$, there have to be non-zero \emph{real} numbers $\alpha,\beta$ such that
\begin{align}
  0=\alpha\d_1C(\theta)+\beta\d_2C(\theta).
\end{align}
Checking all the possible cases of vanishing/non-vanishing sines/cosines, it is easy to see that this condition in unsatisfiable. Hence, both parameters $\theta_1$ and $\theta_2$ are independent at all points in parameter space, and the circuit defines a real two-dimensional submanifold of $\d B_{\cn^2}$.

Of course, in general, checking the solvability of such a set of equations is not efficient if $\theta$ is left a variable. Instead we can write down the Jacobian with real and imaginary parts split;
\begin{align}
  C'(\theta)=\frac12
  \begin{pmatrix}
    -\sin\frac{\theta_1}{2}\cos\frac{\theta_2}{2}&-\cos\frac{\theta_1}{2}\sin\frac{\theta_2}{2}\\
    \cos\frac{\theta_1}{2}\sin\frac{\theta_2}{2}&\sin\frac{\theta_1}{2}\cos\frac{\theta_2}{2}\\
    \sin\frac{\theta_1}{2}\sin\frac{\theta_2}{2}&-\cos\frac{\theta_1}{2}\cos\frac{\theta_2}{2}\\
    -\cos\frac{\theta_1}{2}\cos\frac{\theta_2}{2}&\sin\frac{\theta_1}{2}\sin\frac{\theta_2}{2}
  \end{pmatrix}
  .
\end{align}
Checking this at a some point, e.g., $\theta=0$ for simplicity, we obtain
\begin{align}
  C'(\theta)=\frac12
  \begin{pmatrix}
    0&0\\
    0&0\\
    0&-1\\
    -1&0
  \end{pmatrix}
\end{align}
which is of rank $2$. Thus we conclude that at least in a neighborhood of $\theta=0$ both parameters are independent. The matrix $\tilde C_0(\theta)$ of independent columns near $\theta=0$ therefore coincides with $C'(\theta)$ and $\tilde C_0(\theta)^*\tilde C_0(\theta)$ simplifies to the $\frac14\id_{2\times 2}$ where $\id_{2\times 2}$ is the $2\times2$-identity matrix. This proves that the radius of validity is infinite and therefore that the two parameters $\theta_1$ and $\theta_2$ are always independent.

\subsubsection{Reduction of a four-parameter circuit}\label{sec:4_param_circuit}
As a second example, let us consider the circuit
\begin{align}
  C(\theta)=R_Y(\theta_4)R_X(\theta_3)R_Z(\theta_2)R_X(\theta_1)\ket0.
\end{align}
Then
\begin{align}
  \d_1C(\theta)=\,&\frac{-i}{2}R_Y(\theta_4)R_X(\theta_3)R_Z(\theta_2)XR_X(\theta_1)\ket0,\\
  \d_2C(\theta)=\,&\frac{-i}{2}R_Y(\theta_4)R_X(\theta_3)ZR_Z(\theta_2)R_X(\theta_1)\ket0,\\
  \d_3C(\theta)=\,&\frac{-i}{2}R_Y(\theta_4)XR_X(\theta_3)R_Z(\theta_2)R_X(\theta_1)\ket0,\\
  \d_4C(\theta)=\,&\frac{-i}{2}YR_Y(\theta_4)R_X(\theta_3)R_Z(\theta_2)R_X(\theta_1)\ket0.
\end{align}
If we assume we are using a local optimizer and the procedure starts with $\theta=0$, we observe
\begin{align}
  \d_1C(0)=\,&\frac{-i}{2}X\ket0=\frac{-i}{2}\ket1,\\
  \d_2C(0)=\,&\frac{-i}{2}Z\ket0=\frac{-i}{2}\ket0,\\
  \d_3C(0)=\,&\frac{-i}{2}X\ket0=\frac{-i}{2}\ket1,\\
  \d_4C(0)=\,&\frac{-i}{2}Y\ket0=\frac{1}{2}\ket1.
\end{align}
Of these vectors, the set $\{\d_1C(0),\d_2C(0),\d_4C(0)\}$ is linearly independent over $\rn$. Therefore, we can ignore the parameter $\theta_3$ for this step in the procedure. The submanifold locally defined by the reduced circuit $\check C(\check\theta):=C(\theta)|_{\theta_3=0}$ has dimension $3$, which coincides with $\dim_\rn\d B_{\cn^2}$. The reduced circuit $\check C(\check\theta)$ is therefore maximally expressive at this point.

Considering
\begin{align}
  \check C_1(\check \theta)=R_Y(\check \theta_4)R_Z(\check\theta_2)R_X(\check\theta_1)\ket0,
\end{align}
we know that the Jacobian has rank $3$ near zero. Thus, we can compute $\check C_1'(\check\theta)^*\check C_1'(\check\theta)$ to estimate the area of validity of this reduced circuit.  This computation yields
\begin{align}
  \check C_1'(\check\theta)^*\check C_1'(\check\theta) = 
  \begin{pmatrix}
    \frac14&0&\frac{-\sin\check\theta_2}{4}\\
    0&\frac14&0\\
    \frac{-\sin\check\theta_2}{4}&0&\frac14
  \end{pmatrix}
\end{align}
and therefore 
\begin{align}
  \begin{split}
    \det\l(\check C_1'(\check\theta)^*\check C_1'(\check\theta)\r) =&\, \frac{1-\sin^2\check\theta_2}{64}\\
    =& \,\frac{\cos^2\check\theta_2}{64}.
  \end{split}
\end{align}
The reduced circuit is therefore valid as long as $\cos\check\theta_2\ne0$. Using the estimate on the radius of validity 
\begin{align}
    R\ge\frac{\abs{D(\check\theta)}}{\max_\theta\abs{D'(\theta)}}
\end{align}
with $D(\check\theta):=\det\l(\check C_1'(\check\theta)^*\check C_1'(\check\theta)\r)$, we still obtain the estimate $R\ge1$ for any $\check\theta$ with $\check\theta_2\in\pi\zn$ since $R_0$ can be chosen arbitrarily.

At $\theta_2=\frac\pi2$, we obtain that $\theta_4$ is dependent on $\theta_1$ and $\theta_2$. Hence, we may choose the reduced circuit  
\begin{align}
  \check C_2(\check \theta)=R_X(\check \theta_3)R_Z(\check\theta_2)R_X(\check\theta_1)\ket0,
\end{align}
which now satisfies
\begin{align}
  \begin{split}
    \det\l(\check C_2'(\check\theta)^*\check C_2'(\check\theta)\r) =& \,\frac{1-\sin^2\check\theta_2}{64}\\
    =&\,\frac{\sin^2\check\theta_2}{64}.
  \end{split}
\end{align}
To ensure maximal expressivity, we may therefore consider the circuit 
\begin{align}
  C^+(\theta)=R_Y(\theta_4)R_X(\const)R_Z(\theta_2)R_X(\theta_1)\ket0
\end{align}
whenever $\dist(\theta_2,\pi\zn)\le\frac{\pi}{4}$ and 
\begin{align}
  C^-(\theta)=R_Y(\const)R_X(\theta_3)R_Z(\theta_2)R_X(\theta_1)\ket0
\end{align}
whenever $\dist(\theta_2,\pi\zn)>\frac{\pi}{4}$. Repeating the analysis we performed above for $\const=0$, we observe that this choice between $C^+$ and $C^-$ always provides a valid maximally expressive circuit and the distance from any parameter set $\theta$ to the closest singular point of the restricted circuit $C^\pm$ is $\ge\frac{\pi}{4}$.

\subsubsection{General reduction to finitely many circuits}\label{sec:finitely_many_circuits}
For the particular example given in the previous subsection, none of this elaboration is necessary, but in general we can use this procedure to construct maximally expressive circuits. By starting with a redundant circuit $C$, we can choose a maximally expressive starting restriction $\check C_{\theta_0}$ at an initial $\theta_0$ and compute/estimate its area of validity $O_{\theta_0}$. Choosing a subset $\Omega_{\theta_0}\sse O_{\theta_0}$ (e.g., the interior of a compact) we can ensure that $\Omega_{\theta_0}$ is uniformly bounded away from the set of singular points. If the starting circuit is sufficiently redundant, then each point $\theta$ outside $\Omega_{\theta_0}$ has a maximally expressive restriction $\check C_\theta$ with area of validity $\Omega_\theta$, which, too, is uniformly bounded away from the singular points of $\check C_\theta$. Hence, if the parameter space is compact, there exists finitely many $\theta$ such that the corresponding finitely many $\Omega_\theta$ still form an open cover of the parameter space. Each of these $\Omega_\theta$ is associated with a maximally expressive restriction $\check C_\theta$, and restricting their validity to the $\Omega_\theta$ implies that no singular point can be approximated in parameter space.

In the case above, although we could have restricted to two circuits by setting $\const=0$ in both cases, it seemed more reasonable to still consider an infinite cover because this allows us to perform ``seamless'' switching, i.e., simply change which parameters are treated as constants. We could have chosen the finite cover 
\begin{align}
  C_0^+(\theta)=R_Y(\theta_3)R_Z(\theta_2)R_X(\theta_1)\ket0
\end{align}
whenever $\dist(\theta_2,\pi\zn)\le\frac{\pi}{4}$ and 
\begin{align}
  C_0^-(\theta)=R_X(\theta_3)R_Z(\theta_2)R_X(\theta_1)\ket0
\end{align}
whenever $\dist(\theta_2,\pi\zn)>\frac{\pi}{4}$. However, that would require additional work whenever the algorithm crosses $\theta_2=\frac{\pi}{4}$. In particular, if we are using $C_0^+$ crossing $\theta_2=\frac{\pi}{4}$ to some point $\theta$ and want to switch to $C_0^-$, then we need to compute $\theta_{\rm new}=(C_0^-)^{-1}(C_0^+(\theta))$ to find the new parameter set for the new circuit $C_0^-$. While this is possible (the $C_0^\pm$ are charts of the state space and thus diffeomorphisms), it is much more practical to consider the infinitely many restrictions $C^\pm$ as in the example above.

\subsubsection{Global phase removal}\label{sec:4_param_circuit_global_phase}
Using the analysis techniques for removing symmetries, which we will develop in \autoref{sec:symmetry_removal}, we can further analyze $C^\pm$ and note that the common gates $R_Z(\theta_2)R_X(\theta_1)$ cannot generate an arbitrary phase. This means that the difference between $C^+$ and $C^-$ is solely due to the ability of generating arbitrary phases. Since global phases are irrelevant, we can consider the reduced circuit
\begin{align}
  \check C(\theta):=R_Z(\theta_2)R_X(\theta_1)\ket0. 
\end{align}
This circuit can generate an arbitrary single-qubit state up to a global phase. Using the analysis at the beginning of this current subsection, the circuit is free of singular points.

The same is true for the circuit $R_Z(\theta_2)R_Y(\theta_1)\ket0$. Hence, the first layer of QISKIT's EfficientSU2 2-local circuit \verb|EfficientSU2(3, reps=N)| circuit (cf. \autoref{sec:EfficientSU2}) can generate arbitrary tensor products of single-qubit states (up to a global phase).

\subsection{Return to QISKIT's EfficientSU2 2-local circuit}\label{sec:EfficientSU2_again}
At this point, we can quickly return to the \verb|EfficientSU2(3, reps=N)| and provide the details for the example of \autoref{sec:EfficientSU2} as an illustration. Using the $N=1$ circuit $C(\theta)$ (for simplicity of the output), we can compute the reduced row echelon form for $C'(\theta)$. At $\theta=0$, we obtain 
\begin{widetext}
  \begin{align}
    \left(
    \begin{array}{*{12}c}
      \theta_1&\theta_2&\theta_3&\theta_4&\theta_5&\theta_6&\theta_7&\theta_8&\theta_9&\theta_{10}&\theta_{11}&\theta_{12}\\
      1&0&0&0&0&0&0&0&0&0&0&0\\
      0&1&0&0&0&0&0&0&0&0&0&0\\
      0&0&1&0&0&0&0&0&1&0&0&0\\
      0&0&0&1&1&1&0&0&0&1&1&1\\
      0&0&0&0&0&0&1&0&0&0&0&0\\
      0&0&0&0&0&0&0&1&0&0&0&0\\
      0&0&0&0&0&0&0&0&0&0&0&0\\
      \vdots&\vdots&\vdots&\vdots&\vdots&\vdots&\vdots&\vdots&\vdots&\vdots&\vdots&\vdots      
    \end{array}
    \right)
  \end{align}
\end{widetext}
with the dots indicating that the rest of each column is filled with zeros. Since the $j$\textsuperscript{th} column represents the parameter $\theta_j$, we can identify that parameter $\theta_{9}$ depends on $\theta_3$ and the parameters $\theta_5$, $\theta_6$, $\theta_{10}$, $\theta_{11}$, and $\theta_{12}$ all depend on $\theta_4$. In comparison, the reduced row echelon form of $C'(\theta)$ evaluated at a random parameter set $\theta$ is (with probability $1$) of the form $\begin{pmatrix}A\\B\end{pmatrix}$ where $A$ is the $12\times12$ identity and $B$ the $4\times12$ zero matrix; thus indicating that all twelve parameters are independent.

If, on the other hand, we consider the circuit $\tilde C(\phi,\theta)$, which is leading $C(\theta)$ with $R_Z(\phi)$ on the first qubit, and compute the reduced row echelon form of $\tilde C'(\phi,\theta)$ at $\phi=0$ and $\theta=0$, we obtain
\begin{widetext}
  \begin{align}
    \left(
    \begin{array}{*{13}c}
      \phi&\theta_1&\theta_2&\theta_3&\theta_4&\theta_5&\theta_6&\theta_7&\theta_8&\theta_9&\theta_{10}&\theta_{11}&\theta_{12}\\
      1&0&0&0&1&1&1&0&0&0&1&1&1\\
      0&1&0&0&0&0&0&0&0&0&0&0&0\\
      0&0&1&0&0&0&0&0&0&0&0&0&0\\
      0&0&0&1&0&0&0&0&0&1&0&0&0\\
      0&0&0&0&0&0&0&1&0&0&0&0&0\\
      0&0&0&0&0&0&0&0&1&0&0&0&0\\
      0&0&0&0&0&0&0&0&0&0&0&0&0\\
      \vdots&\vdots&\vdots&\vdots&\vdots&\vdots&\vdots&\vdots&\vdots&\vdots&\vdots&\vdots&\vdots     
    \end{array}
    \right)
  \end{align}
\end{widetext}
with the dots again indicating that the rest of each column is filled with zeros. Once more we can identify $\theta_9$ as dependent on $\theta_3$ but also $\theta_4$, $\theta_5$, $\theta_6$, $\theta_{10}$, $\theta_{11}$, and $\theta_{12}$ are all dependent on $\phi$. Since, by construction, the impact of $\phi$ is only a global phase, we know that the gates corresponding to $\theta_4$, $\theta_5$, $\theta_6$, $\theta_{10}$, $\theta_{11}$, and $\theta_{12}$ only change the phase if varied at $\theta=0$.

In the \verb|EfficientSU2(3, reps=2)| case, the reduced row echelon form of $C'(\theta)$ evaluated at random values of $\theta$ produces
\begin{align}
  \begin{pmatrix}
    \theta_1,\ldots,\theta_{15}&\theta_{16},\theta_{17},\theta_{18}\\
    \id&A\\
    0&0
  \end{pmatrix}
\end{align}
where $A$ is some $15\times3$-matrix. This shows that $\theta_1,\ldots,\theta_{15}$ are independent. Similarly, if we introduce an initial $R_Z(\phi)$ gate on the first qubit and compute the reduced row echelon form for $\tilde C'(\phi,\theta)$ at random values of $(\phi,\theta)$, we obtain
\begin{align}
  \begin{pmatrix}
    \phi&\theta_1,\ldots,\theta_{14}&\theta_{15},\ldots,\theta_{18}\\
    1&0&A\\
    0&\id&B\\
    0&0&0
  \end{pmatrix}
\end{align}
which shows that the net contribution of $\theta_{15}$ is a global phase. Everything else can be generated by $\theta_1,\ldots,\theta_{14}$. It is important to note that although these are pointwise evaluations, sufficiently small perturbations cannot change the set of independent parameters. This is a corollary of the lower-semicontinuity property.

\subsection{Removal of symmetries}\label{sec:symmetry_removal}
As already alluded to at the end of \autoref{sec:EfficientSU2} and \autoref{sec:circuit_reduction_single_qubit}, it can be important to construct a quantum circuit that avoids certain unwanted symmetries, like generating states that differ only by a global phase. Let us more generally consider an $n$-parameter circuit $C(\theta_1,\ldots,\theta_n)$ and suppose that the reduced row echelon form for $C'(\theta)$ is of the form 
\begin{align}\label{eq:circuit_n_independent_params_RREF}
  \begin{pmatrix}
    \id_{n\times n}\\
    0_{(d-n)\times n}
  \end{pmatrix}
\end{align}
where $d$ is the real dimension of the Hilbert space, i.e., $2^{Q+1}$ if $Q$ is the number of qubits. If we wish to avoid an unwanted symmetry that can be encoded using a parametric gate set $U(\phi)$ (e.g., $U(\phi)=R_Z(\phi)$ on some qubit to generate a global phase as in \autoref{sec:EfficientSU2} and \autoref{sec:circuit_reduction_single_qubit}), then instead of analyzing $C$, we may analyze $\tilde C(\phi,\theta)$ given by
\begin{align*}
  \Qcircuit @C=1em @R=.7em {
    \lstick{\ket{0}} & \multigate{2}{U(\phi)} & \multigate{2}{C(\theta)} &\qw\\
    \lstick{\vdots} & \nghost{U(\phi)} & \nghost{C(\theta)} &\vdots\\
    \lstick{\ket{0}} & \ghost{U(\phi)} & \ghost{C(\theta)} &\qw\\
  }
\end{align*}
This encoding $U$ should have at least one parameter value $\phi_0$ for which $U(\phi_0)=\id$. This is necessary for $U$ to ``generate'' the symmetry as opposed to ``changing the circuit to contain the symmetry.'' Furthermore, the dimensionality of the parameter $\phi$ should coincide with the dimension of the unwanted symmetry, e.g., $\phi$ is one-dimensional for the global phase symmetry $U(1)$ or three-dimensional for an $SU(2)$ symmetry.

At this point, it is important to recall the connection between global, local, and pointwise testing of parameter dependence. A circuit is subject to a global symmetry if and only if the action of the symmetry maps the circuit manifold into itself. Given a compact symmetry group, it suffices to check this property locally for the symmetry group. Given a compact circuit manifold, it also suffices to check this property locally for the circuit manifold. The group properties furthermore allow us to only check a neighborhood of the identity element of the symmetry group, as any other pair of open subsets $U$ of the group and $V$ of the circuit manifold can be replaced by $U'=g^{-1}U$ and $V'=g^{-1}V$ where $g\in U$. Then $U'$ is a neighborhood of the identity. Finally, lower-semicontinuity of parameter independence implies that it suffices to test independence on a single point to obtain independence on a neighborhood of the point. Since we plan to use the dimensional expressivity analysis on an extended circuit $\tilde C(\phi,\theta)$, this means that it suffices to check for parameter independence at $\phi_0$ satisfying $U(\phi_0)=\id$ and sufficiently many $\theta$, as discussed in \autoref{sec:semicontinuity}. In this sense, it is sufficient to consider global symmetries, such as the gate $R_Z(\phi)$ which generates a global phase on the current state of the quantum circuit, in order to identify parameters that are responsible for global symmetries.

We can therefore identify the parameters generating a symmetry by looking at parameters that are shown as independent under dimensional expressivity analysis for the circuit $C$ but dependent when analyzing $\tilde C$. For simplicity, let us assume a one-dimensional symmetry. One possible outcome for the reduced row echelon form of $\tilde C'(\phi_0,\theta)$ would be
\begin{align}
  \begin{pmatrix}
    \phi&\theta&\\
    1&0\\
    0&\id_{n\times n}\\
    0_{(d-n-1)\times 1}&0_{(d-n-1)\times n}
  \end{pmatrix}.
\end{align}
If this is the case for almost all parameter sets $\theta$, then all parameters $\theta_1$, $\ldots$, $\theta_n$ are independent and the symmetry encoded by $U$ is not generated by the circuit $C(\theta)$.

The other option is that $C$ does generate the symmetry encoded by $U$. If that is the case, then one\footnote{In general, the number of dependent parameters has to be equal to the dimension of the tested symmetry, i.e., the dimensionality of $\phi$.} of the parameters $\theta_1$, $\ldots$, $\theta_n$ is dependent on $\phi$ and the remaining $\theta_j$. Let this parameter be $\theta_m$. The reduced row echelon form of $\tilde C'(\phi_0,\theta)$ will then be 
\begin{align}\label{eq:theta_m_dependent_RREF}
  \begin{pmatrix}
    \phi&\theta_1,\ldots,\theta_{m-1}&\theta_m&\theta_{m+1},\ldots,\theta_n\\
    1&0&a&0\\
    0&\id&b&0\\
    0&0&0&\id\\
    0&0&0&0
  \end{pmatrix}
\end{align}
where $b$, both $\id$, and each $0$ are matrices/vectors of the appropriate dimension. Crucially, $a$ is non-zero ($a=0$ would violate \autoref{eq:circuit_n_independent_params_RREF}). The vector $b$ may or may not be zero. If $b=0$, then $\theta_m$ has locally the same effect as $U$. The gate corresponding to $\theta_m$ thus locally only contributes the unwanted symmetry. In general, $b$ is expected to be non-zero. In that case, the gate corresponding to $\theta_m$ is not locally equivalent to the unwanted symmetry but removal of the unwanted symmetry from the circuit manifold yields a submanifold that can locally be described by the reduced circuit after removal of the gate corresponding to~$\theta_m$. Thus, avoidance of the unwanted symmetry can locally be achieved by setting $\theta_m$ constant.\footnote{We cannot simply remove the gate $G$ corresponding to $\theta_m$ as this would replace $G(\theta_m)$ with the identity which changes the output of the circuit. If $G=R_X$ and $\theta_m=\pi$ then we would remove an $X$ gate from the circuit and thus fundamentally change the quantum state.}

While it is not clear a priori whether or not the same $\theta_m$ is always responsible for the additional symmetry, it is at least locally always true. In other words, if we find that the $m$\textsuperscript{th} parameter ($\theta_m$) generates an unwanted symmetry at a point $\theta$, then there exists a neighborhood of $\theta$ such that the unwanted symmetry is always generated by the $m$\textsuperscript{th} parameter.

The idea behind the proof is as follows. Since $\theta_m$ generates the unwanted symmetry at the point $\theta$, its partial derivative of the circuit $\d_mC(\theta)$ is a linear combination of the $\phi$ derivative and all other $\theta_j$ derivatives. In other words, we can write 
\begin{align}\label{eq:symmetry_linear_combination}
    \d_mC(\theta)=\alpha_0\d_\phi\tilde C(\phi_0,\theta)+\sum_{j\ne m}\alpha_j\d_jC(\theta) 
\end{align}
for some values $\alpha_j\in \rn$. In order to prove that $\theta_m$ is always responsible for the unwanted symmetry (at least in a neighborhood of $\theta$), then we need to show that we can always find values $\alpha_j$ satisfying \autoref{eq:symmetry_linear_combination} if $\theta$ is subject to sufficiently small perturbations. This can be achieved using the implicit function theorem. We will work out the precise details of this in the remainder of this section.

Since we have $n$ independent vectors $\d_j C(\theta)$, we can assume without loss of generality that we are working in a representation of $\rn^n$. Similarly, we may identify the parameter space (at least locally) with an open subset of $\rn^n$.  Furthermore, without loss of generality let $m=n$. Since $\d_\phi\tilde C(\phi_0,\theta)$ is linearly dependent on the $\d_j C(\theta)$, it too can be represented in $\rn^n$. By \autoref{eq:theta_m_dependent_RREF}, there now exist $\alpha_0,\ldots,\alpha_{n-1}\in\rn$ such that 
\begin{align}
    F(x,y):=\,&\d_nC(x)-y_0\d_\phi\tilde C(\phi_0,x)-\sum_{j=1}^{n-1}y_j\d_jC(x) 
\end{align}
satisfies $F(\theta,\alpha)=0$. Using \autoref{eq:theta_m_dependent_RREF}, we observe that $\d_yF(\theta,\alpha)$ is invertible. The implicit function theorem therefore ensures the existence of a neighborhood $U$ of $\theta$, a neighborhood $V$ of $\alpha$, and a continuously differentiable
$f:\ U\to V$ such that for all $(x,y)\in U\times V$
\begin{align}
  F(x,y)=0\ \iff\ y=f(x). 
\end{align}
In particular, for all $\theta'$ in a neighborhood of $\theta$ there exists $\alpha'$ (in a neighborhood of $\alpha$) such that
\begin{align}
  \d_nC(\theta') = \alpha_0'\d_\phi\tilde C(\phi_0,\theta')+\sum_{j=1}^{n-1}\alpha_j'\d_jC(\theta').
\end{align}
This proves that, at least locally, we can always keep the same parameter constant if we wish to remove an unwanted symmetry from the circuit.

\section{Analysis and design of a translationally invariant $4$-qubit circuit}\label{sec:4Q_circuit_design}
In this section, we will apply the dimensional expressivity analysis to the case of a translationally invariant $4$-qubit circuit. In addition to the question of maximal expressivity, which we focused on in \autoref{sec:EfficientSU2}, the aim of this analysis is to construct a quantum circuit that encodes a particular physical symmetry; in this case, translational invariance. As such, the analysis in this section will heavily depend on \autoref{sec:translation_invariance} in addition to \autoref{sec:parameter_reduction}.

The first circuit under consideration is a circuit that we adapted from a more general framework to study the translationally invariant Ising model, whose Hamiltonian is of the form 
\begin{align*}
  H_{\rm Ising}=J\sum_{q=1}^4X_qX_{q+1}+B\sum_{q=1}^4Z_q,
\end{align*}
where $X_5=X_1$. However, upon integrating the translational invariance, we found the circuit failed to perform.

\begin{figure}[t]  
  \includegraphics[width=\columnwidth]{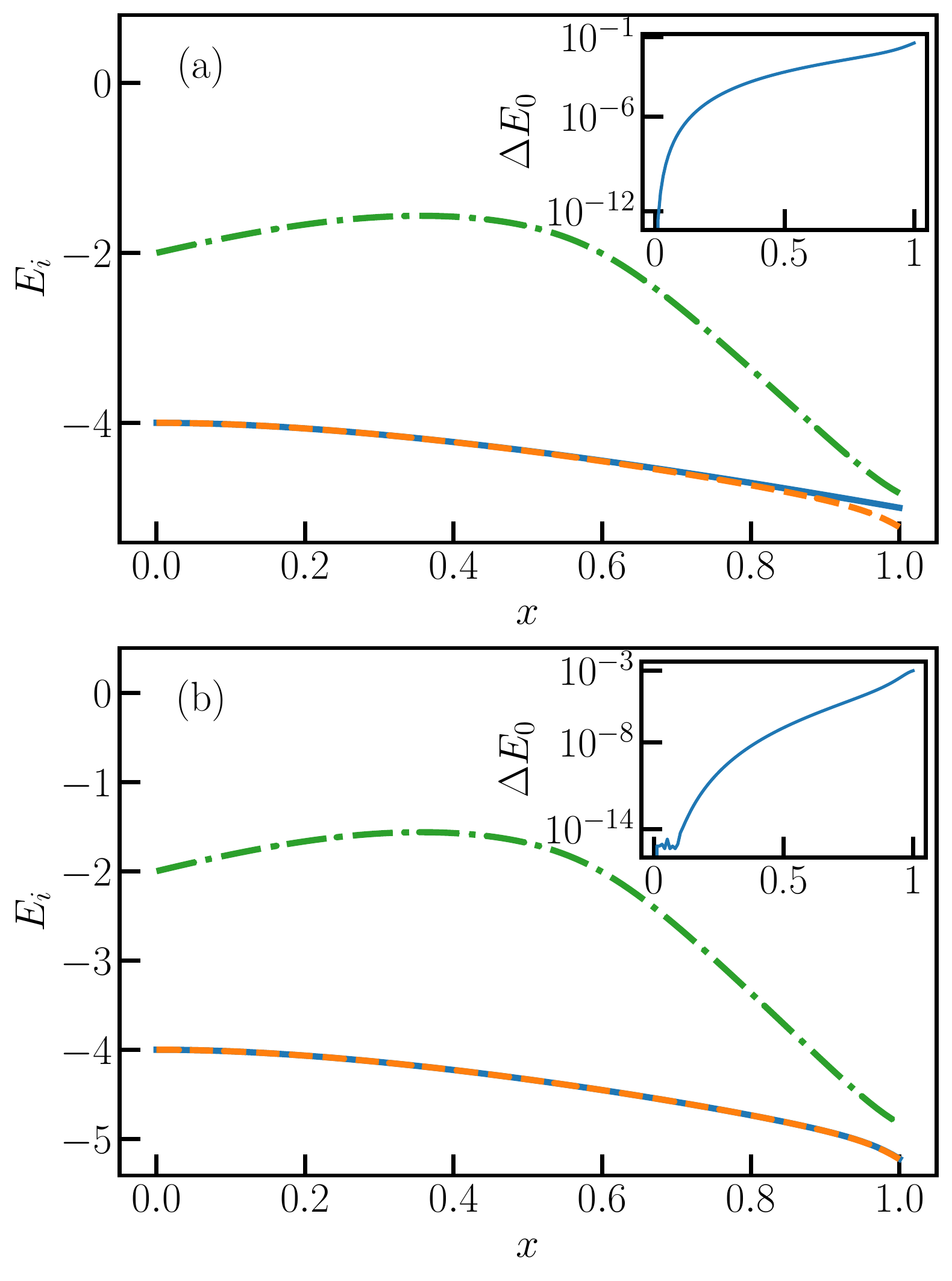}
  \caption{(a) Failure of gradient based variational quantum simulation to continually deform the known ground state of a Hamiltonian $H(0)$ to the target Hamitonian $H(1)$. The VQS ground state energy (blue solid line) deviates from the analytic ground state (orange dashed line). The first excited state energy (green dash-dotted line) is shown as well. The inset shows the relative deviation of the VQS ground-state energy from the exact result, which is of order $\mathcal{O}(0.1)$ for $x\to 1$. (b) Success of the same gradient based variational quantum simulation after adding a different entangler based on expressivity analysis, which reduces the deviation by two orders of magnitude (see inset in the lower panel).}
  \label{fig:Ising}
\end{figure}

The implementation started with a known solution to an instance of the Ising model ($J=0$ and $B=-1$) and then continuously deformed the Hamiltonian from the solved initial Hamiltonian $H_0=-\sum_{q=1}^4Z_q$ to the target Hamiltonian $H_1=\sum_{q=1}^4X_qX_{q+1}-\sum_{q=1}^4Z_q$. Discretizing the path of the Hamiltonian $H(x):=(1-x)H_0+xH_1$ into a large number of steps $H(x_i)$ and using the solution of $H(x_i)$ as initial guess for $H(x_{i+1})$, we aimed to ensure that a local (gradient based) optimizer would not incorrectly terminate due to a local minimum. However, the result is shown in \autoref{fig:Ising}(a). The ground state energy as obtained from a variational quantum simulation visibly deviates from the true ground state energy, although there is a sizable energy gap. This result was independent of the number of layers used for the circuit. Thus, the aim of this example is to use dimensional expressivity analysis to explain why something like this can happen and how to fix it. 

Using a non-translationally invariant circuit that worked well for the more general Heisenberg model, the translational invariance was built into the circuit by replacing each single-qubit rotation gate $R_{O_q}(\theta)=e^{i\frac{\theta}{2}O_q}$, such as $R_{X_q}(\theta)=e^{i\frac{\theta}{2}X_q}$, with a rotation layer
\begin{align}
  L_O(\theta):=\prod_{q=1}^4R_{O_q}(\theta)
\end{align}
and similarly for multi-qubit rotations, e.g.,
\begin{align}
  L_{X_qX_{q+1}}(\theta):=\prod_{q=1}^4R_{X_{q}X_{q+1}}(\theta).
\end{align}
In other words, each gate is translated over all qubits and each translate has the same angle. The circuit in question then has an initial layer
\begin{align}
  L_0:=&\, L_{Z}(\theta_2)L_{X}(\theta_1)
\end{align}
and arbitrarily many further layers of the form
\begin{align}
  \begin{split}
    L_j:=&\, L_{Y}(\theta_{4j+2})L_{Z}(\theta_{4j+1})L_{X}(\theta_{4j})\\
    &\cdot L_{X_{q}X_{q+1}}(\theta_{4j-1})
  \end{split}
\end{align}
where $j\in\{1,\ldots,N\}$ is the layer index. The complete circuit with $N$ layers for the momentum $1$ sector (cf., \autoref{sec:translation_invariance}) is then given by
\begin{align}\label{eq:ising-circuit}
  C(\theta):=L_N\cdots L_1L_0\ket{0}.  
\end{align}
In order to solve the translationally invariant Ising model, we started with an instance of the Ising model whose solution is known analytically (all parameters $\theta_k$ set to $0$) and deformed the model continuously to the target model. This approach was chosen in order to avoid local minima in the energy landscape, which are likely to trap the state starting from a random configuration. Unfortunately, the solution failed to build up the entanglement we expected from the ground state of the final Hamiltonian. In fact, the circuit only produced product states as evidenced by the fact that the angles for the entangling layers $L_{X_qX_{q+1}}$ always remained zero. In other words, the direction of entangled states in the state space, which can be generated using this entangling layer, was orthogonal to the gradient of the energy landscape. Therefore, one of the reasons for the circuit's failure was that -- up to product state optimization -- we were stuck in a local minimum induced by the circuit. 

Using the results of \autoref{sec:parameter_reduction} and \autoref{sec:translation_invariance}, we can analyze this circuit. The initial step of the analysis is to compute the tangent vectors of the circuit with respect to each parameter. These tangent vectors then span the tangent space of the manifold generated by the image of the circuit. Since we started in a known solution, we will exemplify the analysis using the initial parameter set $\theta_{\rm initial}=0$. 

Computing the tangent space of the circuit manifold at $\theta_{\rm initial}=0$ according to \autoref{sec:parameter_reduction} yields that the tangent space is spanned by
\begin{align}
  \begin{split}
    \d_{1}C(0)=\d_{4j}C(0)=&-\frac{i}{2}\sum_qX_q\ket0\\
    \d_2C(0)=\d_{4j+1}C(0)=&-\frac{i}{2}\sum_qZ_q\ket0\\
    \d_{4j+2}C(0)=&-\frac{i}{2}\sum_qY_q\ket0\\
    \d_{4j-1}C(0)=&-\frac{i}{2}\sum_qX_qX_{q+1}\ket0
  \end{split}
\end{align}
for every value of $N$. The circuit therefore defines a $4$-dimensional submanifold in a neighborhood of $\theta=0$, and only the parameters $\theta_1$, $\theta_2$, $\theta_3$, and $\theta_6$ are relevant at the point the circuit failed. This explains why adding additional layers of the same structure has no impact in this case.

Na\"ively, a reasonable idea to overcome this problem would be to replace the $L_{X_qX_{q+1}}$ entangler with a different translationally invariant entangler. However, the natural choices will fail, as well. An additional $L_{Y_qY_{q+1}}$ entangler would generate the direction $\d_3C(0)$ and an additional $L_{Z_qZ_{q+1}}$ would generate $\d_2C(0)$. The na\"ive choices for translationally invariant entanglers in combination with single-qubit rotations cannot generate a circuit manifold of dimension~$>4$. Yet the momentum $1$ sector has dimension $11$ (cf., \autoref{sec:translation_invariance}).

Luckily, our analysis in \autoref{sec:translation_invariance} can guide us in overcoming the impasse. In \autoref{sec:translation_invariance}, we will compute the dimension of the momentum eigensectors. In particular, we will construct a basis of eigenvectors of the momentum sectors by constructing equivalence classes. The construction of these basis states allows us to identify the missing directions from the circuit. Using an orthogonal set of these missing directions $\ket{v_j}$, we can find gates $G_j$ that will generate the missing directions (here $\ket{v_j}=G_j\ket0$), and finally construct one-parameter layers $L_j(\theta_j)=\exp\left(-i\frac{\theta_j}{2}G_j\right)$. Adding these gates to the circuit adds the missing directions $\ket{v_j}$ to the tangent space.

Of course, there is a certain amount of freedom in the construction of these gates $G_j$. In this case, we chose the layers $L_{X_qY_{q+1}}$, $L_{X_qX_{q+2}}$, $L_{X_qY_{q+2}}$, $L_{X_qX_{q+1}X_{q+2}}$, $L_{Y_qY_{q+1}Y_{q+2}}$, $L_{X_qX_{q+1}X_{q+2}X_{q+3}}$, and $L_{X_qX_{q+1}X_{q+2}Y_{q+3}}$. However, we could have chosen $L_{X_qX_{q+1}Y_{q+2}}$ instead of $L_{Y_qY_{q+1}Y_{q+2}}$ to the same effect. In fact, any pair of $X$ gates could be replaces by a pair of $Y$ gates on the same qubits and vice versa.

This yields the (by construction minimal maximally expressive) circuit
\begin{align}
  \begin{split}
    C(\theta):=& \, L_{X_{q}X_{q+1}X_{q+2}Y_{q+3}}(\theta_{11})\\
    &\cdot L_{X_{q}X_{q+1}X_{q+2}X_{q+3}}(\theta_{10})\\
    &\cdot L_{Y_{q}Y_{q+1}Y_{q+2}}(\theta_9)L_{X_{q}X_{q+1}X_{q+2}}(\theta_8)\\
    &\cdot L_{X_qY_{q+2}}(\theta_7)L_{X_qX_{q+2}}(\theta_6)\\
    &\cdot L_{X_qY_{q+1}}(\theta_5)L_{X_qX_{q+1}}(\theta_4)\\
    &\cdot L_Y(\theta_3)L_Z(\theta_2)L_X(\theta_1)\ket0
  \end{split}
\end{align}
and we can repeat the variational quantum simulation of \autoref{fig:Ising}(a) with this new circuit. The results are shown in \autoref{fig:Ising}(b). In particular, we can see that the ground state energy as obtained from the variational quantum simulation is now much closer to the analytic ground state energy, with a relative error that is two orders of magnitude smaller than the one in \autoref{fig:Ising}(a). Thus, this shows that the tailor-made circuit is not subject to the artificial restrictions which caused the original circuit to fail. 

This construction can be generalized to larger numbers of qubits. For example, the two-qubit entanglers $L_{X_qX_{q+1}}$, $L_{X_qY_{q+1}}$, $L_{X_qX_{q+2}}$, $L_{X_qY_{q+2}}$ will no longer be sufficient to generate all two-qubit entangling directions. For more than four qubits, the entangling layers $L_{X_qX_{q+n}}$ and $L_{X_qY_{q+n}}$ for all $n\le\frac{\#\rm{qubits}}{2}$ are required. 

Furthermore, it should be noted that the $L_Y$ layer can be removed from the circuit because it only contributes the ability to generate arbitrary phases (shown in \autoref{sec:circuit_reduction_single_qubit}), i.e., the circuit
\begin{align}
  \begin{split}
    C(\theta):=& \,L_{X_{q}X_{q+1}X_{q+2}Y_{q+3}}(\theta_{10})\\
    &\cdot L_{X_{q}X_{q+1}X_{q+2}X_{q+3}}(\theta_9)\\
    &\cdot L_{Y_{q}Y_{q+1}Y_{q+2}}(\theta_8)L_{X_{q}X_{q+1}X_{q+2}}(\theta_7)\\
    &\cdot L_{X_qY_{q+2}}(\theta_6)L_{X_qX_{q+2}}(\theta_5)\\
    &\cdot L_{X_qY_{q+1}}(\theta_4)L_{X_qX_{q+1}}(\theta_3)\\
    &\cdot L_Z(\theta_2)L_X(\theta_1)\ket0
  \end{split}\label{eq:ti_ising_example_maximally_expressive}  
\end{align}
can generate arbitrary states with momentum~$1$ (cf., \autoref{sec:translation_invariance}) up to a phase. By construction of the quantum simulation, we only needed to find a custom circuit for the momentum~$1$ sector. This is because having a non-vanishing energy gap throughout the entire quantum simulation implies a continuous deformation of the ground state from the known initial ground state to the target Hamiltonian ground state. In general, it is unlikely that prior knowledge of the system will allow us to know which momentum sector the ground state is in. The process highlighted in this section would therefore need to be repeated for each momentum sector. Applying the tools developed in \autoref{sec:translation_invariance} to the other momentum sectors, will yield custom circuits for each. In fact, the here considered case is the worst case scenario because all other momentum sectors need at most $8$-parameter quantum circuits (\autoref{eq:eigenspace_dimension+2} and removal of a global phase). The explicit computation in \autoref{sec:translational_invariance_Q4} indeed shows that the momentum~$-1$ and momentum~$\pm i$ sectors can be parametrized (up to a phase) with $6$-parameter and $4$-parameter circuits, respectively. Compared to the $30$-parameter circuit that a na\"ive $4$-qubit quantum simulation requires, this shows the power of custom circuit design based on dimensional expressivity analysis.
 
From a practical point of view, the three-qubit interactions in \autoref{eq:ti_ising_example_maximally_expressive} are in general harder to realize than the two-qubit interactions in the original ansatz in \autoref{eq:ising-circuit}. In particular, taking into account the restricted native gate set of current quantum hardware, synthesizing the three-qubit terms can lead to deeper circuits and, thus, to an increase in noise. As a result, on NISQ devices there might be a trade-off between maximal expressivity and simplicity of the ansatz. For example, using only single-qubit rotation layers and all two-qubit rotation layers is sufficient to overcome the above-mentioned failure of an initial circuit being stuck in a product state optimization loop. In this sense, maximal expressivity can be traded for a simpler set of gates, and dimensional expressivity analysis allows us to quantify this gate set deficiency.

\section{Translationally invariant circuits}\label{sec:translation_invariance}
The circuit parameter reduction method described in \autoref{sec:parameter_reduction} is of particular importance if we aim to incorporate (rather than remove) problem-specific symmetries into the quantum circuit as well. These symmetries often partition the search space into physically relevant sectors, which can be searched separately. As such, a quantum circuit that automatically satisfies the symmetries can greatly improve the performance of a quantum simulation, because it only searches the physically relevant subspace of the state space. However, since the measure of deficiency is the codimension of the manifold of reachable states within each physical sector, it is imperative to compute the dimension of these sectors. 

In this section, we will explicitly treat the case of translational invariance. This provides an explicit example that can be used to treat other symmetries analogously. Furthermore, translational invariance is a common property in local Hamiltonians. In this case, the physical sectors are the intersections of the unit sphere of the Hilbert space with the eigenspaces of the translation operator. In \autoref{sec:translationa_invariance_general_case}, we will construct the physical sectors and compute their dimension. As such, we will prove the following \autoref{thm:eigenspace_dimension}. In particular, the real dimension of the physical sectors can then be compared to the circuit manifold dimension as obtained in \autoref{sec:parameter_reduction}, in order to compute the codimension deficiency.

\begin{theorem}\label{thm:eigenspace_dimension}
  Let $Q$ be the number of qubits, $\omega$ be a root of unity of order $d$, i.e., $\omega^d=1$ and $\fa n\in\nn_{<d}:\ \omega^n\ne1$. If $d\mid Q$ (``$d$ divides $Q$''), then $\omega$ is an eigenvalue of the translation operator.

  Furthermore, let $\#(d):=2^d-\sum_{\substack{d'\mid d\\d'<d}}\#(d')$ with $\#(1)=2$ and $[\#](d)=\frac{\#(d)}{d}$. Then, the complex dimension $\dim_\cn\Eig(\omega)$ of the eigenspace $\Eig(\omega)$ is given by
  \begin{align}
    \dim_\cn\Eig(\omega)=&\sum_{d\mid k\mid Q}[\#](k).
  \end{align}
  The real dimension of the physical sector corresponding to $\omega$ is therefore $2\dim_\cn\Eig(\omega)-1$. 
\end{theorem}

Furthermore, we will show in \aref{app:eigenspaces_path_connected} that the physical sector corresponding to each root of unity $\omega$ is path connected. Thus, given $\omega$, any two states in the physical sector corresponding to $\omega$, and any maximally expressive circuit for the physical sector, there exists a path in parameter space connecting the two states. This is particularly important for variational quantum simulations, as it implies that the optimal state can be reached by a path from arbitrary initial states. However, since it is not directly relevant for the circuit analysis, we will provide the proof and discussion of path connectedness in the appendix.

To illustrate the construction and \autoref{thm:eigenspace_dimension}, we will consider the $4$-qubit case in \autoref{sec:translational_invariance_Q4}. We will apply these results to the analysis and design of a $4$-qubit translationally invariant circuit for the momentum $1$ sector in \autoref{sec:4Q_circuit_design}.

\subsection{The general case}\label{sec:translationa_invariance_general_case}
Let us consider a $Q$-qubit system. A basis of the Hilbert space is given by $B:=\{\ket j;\ j\in \nn_{0,<2^Q}\}$, where $\ket j$ denotes the tensor product state $\ket{b_Q(j)_{Q-1}\ldots b_Q(j)_0}$ if and only if $b_Q(j):=b_Q(j)_{Q-1}\ldots b_Q(j)_0$ is the binary representation of $j$. For instance, $b_3(5)=101$ with $b_3(5)_2=b_3(5)_0=1$ and $b_3(5)_1=0$.

The translation operator $\tau_Q$ is defined as the linear operator mapping $B$ to itself and satisfying 
\begin{align}
  b_Q(\tau_Q(j)):=b_Q(j)_{Q-2}\ldots b_Q(j)_0b_Q(j)_{Q-1}
\end{align}
for every $j\in\nn_{0,<2^Q}$. The set of possible momenta is then given by the set of eigenvalues of~$\tau_Q$ and the momentum sectors are the eigenspaces of~$\tau_Q$. In order to compute the dimension of each momentum sector, we define the following equivalence relation on $\nn_{0,<2^Q}$:
\begin{align}
  j\sim_Qk\ :\iff\ \ex t\in\nn:\ j = \tau_Q^t(k).
\end{align}
For example, we observe
\begin{itemize}
\item $1\sim_22$ since $01=\tau_2(10)$ in binary,
\item $5\sim_33$ since $101=\tau_3^2(011)$ in binary,
\item $\fa j\in \nn_{0,<2^Q}:\ j\sim_Qj$ since $j=\tau_Q^Q(j)$.
\end{itemize}
We denote the equivalence class of $j$ in $\nn_{0,<2^Q}/\sim_Q$ by $[j]_Q$. Then we define the order of $j$ to be 
\begin{align}
  \ord_Q(j):=\min\{t\in\nn;\ \tau_Q^t(j)=j\}.
\end{align}
By Lagrange's theorem, we know 
\begin{align}
  \fa j\in \nn_{0,<2^Q}:\ \ord_Q(j)\text{ divides }Q,
\end{align}
e.g., we observe $\ord_4(8)=2$ and $[8]_4=\{8,5\}$ since $b_4(8)=1010$ and $b_4(5)=0101$.

Going forward, it will be advantageous to choose a canonical ordering of the elements in~$[j]_Q$. For simplicity, we will order them by value, that is 
\begin{align}
  [j]_Q=\{j_0,\ldots,j_{\ord_Q(j)-1}\}
\end{align}
where $m<n$ implies $j_m<j_n$, i.e., $[8]_4=\{8_0,8_1\}$ with $8_0=5$ and $8_1=8$.
We can use this ordering to define
\begin{align}
  \omega_Q(j_n):=\exp\l(2\pi i\frac{n}{\ord_Q(j_n)}\r)
\end{align}
which is a bijection between the elements of $[j]_Q$ and the roots of unity of order $\ord_Q(j)$.
This implies that we can construct an eigenvector $e_Q(j_n)$ of $\tau_Q$ for each element of each $[j]_Q$ by setting
\begin{align}
  \tilde e_Q(j_n):=\sum_{k=0}^{\ord_Q(j)-1}\omega_Q(j_n)^k\ket{\tau_Q^k(j_0)}
\end{align}
and
\begin{align}
  e_Q(j_n):=\frac{\tilde e_Q(j_n)}{\norm{\tilde e_Q(j_n)}_{\ell_2(Q)}}.
\end{align}
For example, $[1]_2=\{1,2\}$ implies $e_2(1)=\frac{\ket{01}+\ket{10}}{\sqrt2}$ and $e_2(2)=\frac{\ket{01}-\ket{10}}{\sqrt2}$.

By construction, the $e_Q(j_n)$ are independent of all $e_Q(k_m)$ if $[j]_Q\ne[k]_Q$ because they are linear combinations of linearly independent sets. Furthermore, the $e_Q(j_n)$ are independent of the remaining $e_Q(j_m)$ since they are eigenvectors of different eigenvalues of $\tau_Q$. In other words, we have found a basis transformation between $B$ and a basis of eigenvectors of $\tau_Q$. In particular, we have found a complete set of eigenvectors for $\tau_Q$.

Additionally, we can pigeon hole the number of eigenvectors to a given eigenvalue~$\omega$. Let $\omega$ be a root of unity of order $N$, i.e., $\omega^N=1$ and $\fa n\in\nn_{<N}:\ \omega^n\ne1$. Then, the number $\nu(\omega,j)$ of eigenvectors corresponding to $\omega$ generated from elements in $[j]_Q$ is
\begin{align}
  \nu(\omega,j)=
  \begin{cases}
    1&,\ N\mid\ord_Q(j),\\
    0&,\ N\nmid\ord_Q(j),
  \end{cases}
\end{align}
where $a\mid b$ denotes ``$a$ divides $b$''. Hence, the real dimension of the eigenspace $\Eig(\omega)$ of $\tau_Q$ with respect to the eigenvalue $\omega$ is given by
\begin{align}
  \dim_\cn\Eig(\omega)=\#\{[j]_Q;\ N\mid\ord_Q(j)\}.
\end{align}
In particular, since $1$ divides every $\ord_Q(j)$, we directly conclude
\begin{align}
  \dim_\cn\Eig(1)=\#\{[j]_Q;\ j\in\nn_{0,<2^Q}\}
\end{align}
and 
\begin{align}
  \dim_\cn\Eig(1)\ge\dim_\cn\Eig(\omega)
\end{align}
for every eigenvalue $\omega$ of $\tau_Q$. In fact, since $\ket{0\ldots0}$ and $\ket{1\ldots1}$ are only in $\Eig(1)$ and every generating $j_k$ in $\Eig(\omega)$ is also in $\Eig(1)$, we obtain
\begin{align}\label{eq:eigenspace_dimension+2}
  \dim_\cn\Eig(1)\ge\dim_\cn\Eig(\omega)+2.
\end{align}
It should be noted that \autoref{eq:eigenspace_dimension+2} is optimal. If $Q$ is a power of $2$, then all $\ord_Q(j)$ have to be even or~$1$. Thus, every $j$ with $\ord_Q(j)>1$ has to give an element of $\Eig(1)$ and $\Eig(-1)$ with only $\ket{0\ldots0}$ and $\ket{1\ldots1}$ having order $1$, i.e.,
\begin{align}
  \dim_\cn\Eig(1)=\dim_\cn\Eig(-1)+2.
\end{align}
To compute the dimension of each momentum sector explicitly, let $1=d_1<d_2<\ldots<d_N=Q$ be the divisors of $Q$. Then, each possible order has to be one of the $d_n$. Furthermore, for each $d_n$ there exists a $j\in \nn_{0,<2^Q}$ with $\ord_Q(j)=d_n$ since, setting $r_n:=\frac{Q}{d_n}$, we can define
\begin{align}
  \begin{split}
    j:=&\sum_{k=0}^{r_n-1}2^{kd_n}\\
    =&\overbrace{\underbrace{0\ldots0}_{d_n-1}1\ \underbrace{0\ldots0}_{d_n-1}1\ \ldots\ \underbrace{0\ldots0}_{d_n-1}1}^{r_n\text{blocks of size }d_n}
  \end{split}
\end{align}
to obtain an element $j$ of order $d_n$. Hence, for every divisor $d$ of $Q$ and for every root of unity~$\omega$ of order $d$, there exists a state of momentum $\omega$. This also allows us to count the number of elements $j\in\nn_{0,<2^Q}$ which have order $d$ since their binary representation is built up of $\frac{Q}{d}$ copies of binary strings of length $d$. There are $2^d$ such binary strings, however this is counting all $j$ with $\ord_Q(j)\mid d$ as well. Hence,
\begin{align}\label{eq:divisor_count}
  \begin{split}
    \#(d):=\,&\#\{j;\ \ord_Q(j)=d\}\\
    =\,&2^d-\sum_{\substack{d'\mid d\\d'<d}}\#(d')
  \end{split}
\end{align}
gives a recursion with $\#(1)=2$.

Since the elements of $\nn_{0,<2^Q}$ which are of order~$d$ are grouped into equivalence classes of $d$ elements each, we directly obtain that the number of equivalence classes $[\#](d)$ of order $d$ is given by
\begin{align}\label{eq:reduced_divisor_count}
  [\#](d)=\frac{\#(d)}{d}.
\end{align}
Hence, we obtain 
\begin{align}
  \begin{split}
    \dim_\cn\Eig(\omega)=&\#\{[j]_Q;\ d\mid\ord_Q(j)\}\\
    =&\sum_{d\mid k\mid Q}[\#](k)
  \end{split}
\end{align}
which completes the proof of \autoref{thm:eigenspace_dimension}.

\subsection{Example: the $4$-qubit case}\label{sec:translational_invariance_Q4}
In this section, we will consider the $Q=4$-qubit case as a fully worked example. The construction of the eigenspaces of the 4-qubit translation operator is based on the construction of a suitable mapping between vectors in the 4-qubit Hilbert space $\cn^{16}$ and translational equivalence classes of the computational basis of a 4-qubit quantum device. The equivalence classes $[j]_4$ with respect to
\begin{align}
  j\sim_4k\ :\iff\ \ex t\in\nn:\ j = \tau_4^t(k)
\end{align}
are
\begin{itemize}
\item $[0]_4=\{0=0000\}$,
\item $[1]_4=\{1=0001,2=0010,4=0100,8=1000\}$,
\item $[3]_4=\{3=0011,6=0110,12=1100,9=1001\}$,
\item $[5]_4=\{5=0101,10=1010\}$,
\item $[7]_4=\{7=0111,14=1110,13=1101,11=1011\}$,
\item $[15]_4=\{15=1111\}$,
\end{itemize}
for which we observe the orders $\ord_4(0)=1$, $\ord_4(1)=4$, $\ord_4(3)=4$, $\ord_4(5)=2$, $\ord_4(7)=4$, and $\ord_4(15)=1$.

Each eigenspace corresponds to a root of unity whose order divides $4$ since these are precisely the eigenvalues of the translation operator on 4 qubits. Furthermore, the equivalence class $[j]_4$ contributes exactly one basis vector for the eigenspace $\Eig(\omega)$ corresponding to the root of unity $\omega$ if and only if the order of $j$ is divisible by the order of $\omega$. This implies that we can obtain the dimension of each eigenspace by counting.
\begin{itemize}
\item $\dim\Eig(1)=\#\{[0]_4$, $[1]_4$, $[3]_4$, $[5]_4$, $[7]_4$, $[15]_4\}=6$ (orders divisible by $1$),
\item $\dim\Eig(-1)=\#\{[1]_4,[3]_4,[5]_4,[7]_4\}=4$ (orders divisible by $2$),
\item $\dim\Eig(i)=\#\{[1]_4,[3]_4,[7]_4\}=3$ (orders divisible by $4$),
\item $\dim\Eig(-i)=\#\{[1]_4,[3]_4,[7]_4\}=3$ (orders divisible by $4$).
\end{itemize}
Furthermore, the construction of \autoref{sec:translationa_invariance_general_case} explicitly allows us to identify a basis for each eigenspace. 
\begin{itemize}
\item $\Eig(1)$:
  \begin{enumerate}
  \item $e_4(0_0)=\ket{0000}$,
  \item $e_4(1_0)=\frac{\ket{0001}+\ket{0010}+\ket{0100}+\ket{1000}}{\sqrt4}$,
  \item $e_4(3_0)=\frac{\ket{0011}+\ket{0110}+\ket{1100}+\ket{1001}}{\sqrt4}$,
  \item $e_4(5_0)=\frac{\ket{0101}+\ket{1010}}{\sqrt2}$,
  \item $e_4(7_0)=\frac{\ket{0111}+\ket{1110}+\ket{1101}+\ket{1011}}{\sqrt4}$,
  \item $e_4(15_0)=\ket{1111}$.
  \end{enumerate}
\item $\Eig(-1)$:
  \begin{enumerate}
  \item $e_4(1_1)=\frac{\ket{0001}-\ket{0010}+\ket{0100}-\ket{1000}}{\sqrt4}$,
  \item $e_4(3_1)=\frac{\ket{0011}-\ket{0110}+\ket{1100}-\ket{1001}}{\sqrt4}$,
  \item $e_4(5_1)=\frac{\ket{0101}-\ket{1010}}{\sqrt2}$,
  \item $e_4(7_1)=\frac{\ket{0111}-\ket{1110}+\ket{1101}-\ket{1011}}{\sqrt4}$.
  \end{enumerate}
\item $\Eig(i)$: 
  \begin{enumerate}
  \item $e_4(1_2)=\frac{\ket{0001}+i\ket{0010}-\ket{0100}-i\ket{1000}}{\sqrt4}$,
  \item $e_4(3_2)=\frac{\ket{0011}+i\ket{0110}-\ket{1100}-i\ket{1001}}{\sqrt4}$,
  \item $e_4(7_2)=\frac{\ket{0111}+i\ket{1110}-\ket{1101}-i\ket{1011}}{\sqrt4}$.
  \end{enumerate}
\item $\Eig(-i)$:
  \begin{enumerate}
  \item $e_4(1_3)=\frac{\ket{0001}-i\ket{0010}-\ket{0100}+i\ket{1000}}{\sqrt4}$,
  \item $e_4(3_3)=\frac{\ket{0011}-i\ket{0110}-\ket{1100}+i\ket{1001}}{\sqrt4}$,
  \item $e_4(7_3)=\frac{\ket{0111}-i\ket{1110}-\ket{1101}+i\ket{1011}}{\sqrt4}$.
  \end{enumerate}  
\end{itemize}
Finally, we can compare these explicit results to the abstract results of \autoref{sec:translationa_invariance_general_case}. Using \autoref{eq:divisor_count} and \autoref{eq:reduced_divisor_count}, we obtain
\begin{itemize}
\item $\#(1)=2$ ($0$ and $15$),
\item $\#(2)=2^2-2=2$ ($5$ and $10$),
\item $\#(4)=2^4-2-2=12$ (everything else).
\end{itemize}
as well as
\begin{itemize}
\item $[\#](1)=2$ (corresponding to $[0]_4$ and $[15]_4$),
\item $[\#](2)=\frac{\#(2)}{2}=1$ (corresponding to $[5]_4$),
\item $[\#](4)=\frac{\#(4)}{4}=3$ (corresponding to $[1]_4$, $[3]_4$, and $[7]_4$).
\end{itemize}
and thus with \autoref{thm:eigenspace_dimension}
\begin{itemize}
\item $\dim_\cn\Eig(1)=[\#](1)+[\#](2)+[\#](4)=2+1+3=6$,
\item $\dim_\cn\Eig(-1)=[\#](2)+[\#](4)=1+3=4$,
\item $\dim_\cn\Eig(i)=[\#](4)=3$,
\item $\dim_\cn\Eig(-i)=[\#](4)=3$.
\end{itemize}
As such, we can see that the eigenspaces of the momentum operator have the correct dimensions (summing to $\dim_\cn\cn^{2^4}=16$). Furthermore, the inductive procedure of \autoref{thm:eigenspace_dimension} enables us to quickly identify the required number of parameters for a maximally expressive quantum circuit for each of these momentum sectors. As a corollary, we also obtain structural information about the momentum sectors, which can be used in custom designing quantum circuits similar to the example provided in \autoref{sec:4Q_circuit_design}.

\section{Hardware efficient implementation}\label{sec:implementation}

In this section, we will discuss possible hardware efficient implementation and automation for the proposed dimensional expressivity analysis. Since the analysis requires testing the Jacobian of the quantum circuit for its rank, a classical implementation requires exponential memory in the number of qubits $Q$ (the Jacobian is a matrix of dimension $(\#\text{parameters})\times 2^{Q+1}$). We are therefore looking for either a pure quantum algorithm or a hybrid quantum-classical algorithm that can reduce the classical part of the algorithm to scale polynomially in the number of parameters. Since the dimensional expressivity analysis is performed for each parameter at a time, it requires at least polynomial cost in the number of parameters. As such, ``efficiency'' is also related to the scaling factor between the number of qubits $Q$ and the dimension of the physical state space for a given number of qubits.\footnote{For many physically interesting examples, such as local Hamiltonians, polynomial scaling of the state space dimension in $Q$ implies polynomial scaling of the dimensional expressivity analysis in $Q$ as well. However, if the state space grows exponentially in $Q$, then the dimensional expressivity analysis will also scale exponentially if the entire state space needs to be parametrized.}

In \autoref{sec:hybrid-algorithm}, we will derive a hybrid quantum classical algorithm that efficiently performs the dimensional expressivity analysis at a given set of parameters. As such, it can be used for circuit design if tested on sufficiently many parameter sets, as well as for on the fly circuit reduction, e.g., to only update relevant parameters in a variational quantum simulation. In \autoref{sec:automated-termination}, we will briefly discuss the termination condition for the automated dimensional expressivity analysis as well as error estimates. This ensures that even if a generic scheme producing quantum circuits with arbitrarily many parameters is used, the automated procedure will reject any additional parameters. Finally, in \autoref{sec:hardware-implementation}, we will discuss hardware implementations on IBMQ. 

\subsection{A hybrid quantum-classical algorithm for dimensional expressivity analysis}\label{sec:hybrid-algorithm}
Given a quantum circuit $C$ with $N$ parameters, we recall that the real Jacobian $J$ is given by
\begin{align}
  J=
  \begin{pmatrix}
    \mid&&\mid\\
    \Re\d_1C&\cdots&\Re\d_NC\\
    \mid&&\mid\\
    \\
    \mid&&\mid\\
    \Im\d_1C&\cdots&\Im\d_NC\\
    \mid&&\mid\\
  \end{pmatrix}.
\end{align}
We will perform the dimensional expressivity analysis for each parameter at a time starting with the first. In other words, we will start checking whether the second parameter is independent of the first. Then, we test the third parameter against the first and second, and continue checking each new parameter against its predecessors. Thus, in order to check the $k$\textsuperscript{th} parameter, we need consider the reduced Jacobian
\begin{align}
  \begin{split}
    J_k=&
    \begin{pmatrix}
      \mid&&\mid\\
      \Re\d_1C&\cdots&\Re\d_kC\\
      \mid&&\mid\\
      \\
      \mid&&\mid\\
      \Im\d_1C&\cdots&\Im\d_kC\\
      \mid&&\mid\\
    \end{pmatrix}\\
    =&
    \begin{pmatrix}
      \mid&\mid\\
      \mid&\Re\d_kC\\
      \mid&\mid\\
      J_{k-1}\\
      \mid&\mid\\
      \mid&\Im\d_kC\\
      \mid&\mid\\
    \end{pmatrix},
  \end{split}
\end{align}
where we may assume that $J_{k-1}$ has full rank (because we will have removed any prior dependent parameters). $J_1$ is always full rank, as long as the first parameter does not correspond to a trivial gate. We assume that checking non-triviality is at least possible for the first gate with respect to the ordering of parameters. Any trivial gates corresponding to later parameters will be found by the dimensional expressivity analysis.

Since we are unaware of a quantum algorithm that can check linear independence or the rank of a given dense real matrix, we expect to have to test the rank of each $J_k$ classically. In order to remove exponential memory scaling, we note that 
\begin{align}
  S_k:=J_k^*J_k
\end{align}
is a real $k\times k$-matrix with the same rank as $J_k$. In particular, assuming $J_{k-1}$ has full rank, $S_k$ can have at most a rank deficiency of $1$. Hence, $S_k$ is invertible if and only if the new parameter is independent. Invertibility of $S_k$ can be checked in various ways, e.g., by computing its determinant or its eigenvalues. In either case, since $S_k$ is a square matrix of size $k$, this operation can be performed classically at a cost of $O(k^3)$. Hence, checking invertibility for all $S_k$ with $1\le k\le N$ has a cost of $O(N^4)$, independent of the number of qubits.

Of course, this requires efficient access to the matrices $S_k$. Considering the $(m,n)$-entry $(S_k)_{m,n}$ of $S_k$, we note
\begin{align}
  \begin{split}
    (S_k)_{m,n} =\,& \Re\d_mC^*\Re\d_nC+\Im\d_mC^*\Im\d_nC\\
    =\,&\Re\l\langle\d_mC,\d_nC\r\rangle.
  \end{split}
\end{align}
In other words, we need to efficiently estimate the $\Re\l\langle\d_mC,\d_nC\r\rangle$. Following~\cite{Zhao-Zhao-Rebentrost-Fitzsimmons2019,Lloyd-Mohseni-Rebentrost2013}, this is generally possible using quantum RAM calls. However, for many applications in variational quantum simulations, a more direct approach is possible. 

To efficiently compute the matrices $S_k$, we will restrict the type of parametric gates to be rotation gates of the form 
\begin{align}
  R_G(\theta):=\exp\l(-\frac{i}{2}\theta G\r),
\end{align}
where $G$ is a gate. This class of rotation gates includes the common $R_X$, $R_Y$, and $R_Z$ gates but may also include entangling gates such as $R_{\CNOT}$, $R_{X_qX_{q'}}$, or even more complicated gates $R_G$ with $G=\sum_qX_qX_{q+1}X_{q+2}Y_{q+3}$ as discussed in \autoref{sec:4Q_circuit_design} (i.e., gates necessary to enforce physical symmetries).

Using a circuit $C$ that comprises only gates $R_G$ as described above and non-parametric gates, we can write 
\begin{align}
  \Re\l\langle\d_mC,\d_nC\r\rangle=\frac{1}{4}\Re\l\langle\gamma_m,\gamma_n\r\rangle ,
\end{align}
where each $\gamma_j$ coincides with $C$ except for an additional gate $G_j$ inserted after the application of $R_{G_j}(\theta_j)$. For example, if 
\begin{align}
  C(\theta_1,\theta_2)=R_Z(\theta_2)R_X(\theta_1)\ket0,
\end{align}
then
\begin{align}
  \gamma_1=R_Z(\theta_2)XR_X(\theta_1)\ket0
\end{align}
and
\begin{align}
  \gamma_2=ZR_Z(\theta_2)R_X(\theta_1)\ket0.
\end{align}
In particular, we note that for all such circuits $C$, the diagonal elements of $S_k$ are given by
\begin{align}
  \Re\l\langle\d_nC,\d_nC\r\rangle=\frac{1}{4}.
\end{align}

In order to compute $\Re\l\langle\gamma_m,\gamma_n\r\rangle$ with $m\ne n$ on a quantum device, we need to construct the state~\cite{Zhao-Zhao-Rebentrost-Fitzsimmons2019}
\begin{align}
  \ket{\phi_{m,n}}=\frac{\ket0\otimes\ket{\gamma_m}+\ket1\otimes\ket{\gamma_n}}{\sqrt{2}}.
\end{align}
Using the setup and circuit assumptions laid out above, this requires an ancillary qubit and insertion of controlled gates $CG_j$ after $R_{G_j}$ where $CG_m$ is controlled by the ancilla being $\ket0$ and $CG_n$ is controlled by the ancilla being $\ket1$. The former can be achieved using a standard ``control if $\ket1$'' gate $CG_j$ if it is conjugated with $X$ gates on the ancilla, i.e., inserting $X_\anc CG_j X_\anc$ after $R_{G_j}$. For example, the state $\ket{\phi_{2,1}}$ for
\begin{align}
  C(\theta_1,\theta_2)=R_Z(\theta_2)R_X(\theta_1)\ket0
  \label{eq:example_circuit}
\end{align}
is obtained using the following circuit:
\begin{align*}
  \Qcircuit @C=1em @R=.7em {
    \lstick{\ket{0}} & \gate{R_X(\theta_1)} & \targ    & \gate{R_Z(\theta_2)} & \ctrl{0} &\qw        & \qw\\
    \lstick{\ket{0}} & \gate{H} & \ctrl{-1} & \gate{X} & \ctrl{-1} & \gate{X} &\qw\\
  }
\end{align*}

\begin{figure*}[!ht]
  \begin{align*}
    \Qcircuit @C=1em @R=.7em {
      \lstick{\ket{0}} & \gate{R_X(\theta_1)} & \targ & \gate{R_Z(\theta_2)} & \ctrl{0} &\qw        & \qw       & \qw & \qw\\
      \lstick{\ket{0}} & \gate{H}                    &  \ctrl{-1} & \gate{X}                    & \ctrl{-1} & \gate{X} &\gate{H} & \meter &\cw\\
    }
  \end{align*}
  \caption{Circuit to compute $\Re\l\langle\d_2C,\d_1C\r\rangle$ using an ancilla qubit (lower quantum wire) for the circuit in \autoref{eq:example_circuit}.} \label{fig:DEA-circuit}
\end{figure*}

Once the state $\ket{\phi_{m,n}}$ is prepared, we can apply a final Hadamard gate on the ancilla. This yields the state
\begin{align}
  \begin{split}
    &H_\anc\ket{\phi_{m,n}}\\
    =&\,\frac{\ket0\otimes\l(\ket{\gamma_m}+\ket{\gamma_n}\r)+\ket1\otimes\l(\ket{\gamma_m}-\ket{\gamma_n}\r)}{2}
  \end{split}
\end{align}
and measuring the ancilla yields 
\begin{align}
  \begin{split}
    \prob(\anc=0)=&\,\frac{\l(\bra{\gamma_m}+\bra{\gamma_n}\r)\l(\ket{\gamma_m}+\ket{\gamma_n}\r)}{4}\\
    =&\,\frac{1+\Re\l\langle\gamma_m,\gamma_n\r\rangle}{2}
  \end{split}
  \label{eq:real_part_overlap}
\end{align}
for the ancilla being measured in $\ket0$. 

Returning to the $C=R_ZR_X\ket0$ example, we thus obtain
\begin{align}
  \Re\l\langle\gamma_2,\gamma_1\r\rangle=2\prob(\anc=0)-1,
\end{align}
retrieving $\prob(\anc=0)$ from the circuit shown in \autoref{fig:DEA-circuit}.

In general, this means that we can compute all of the matrices $S_k$ ($1\le k\le N$) using $O(N^2\eps^{-2})$ quantum device calls. Here, $\eps$ is the acceptable error for the estimates of $\Re\langle\gamma_m,\gamma_n\rangle$; e.g., a suitable multiple of the maximal standard deviation. The complete cost of the hybrid quantum-classical algorithm is therefore in $O(N^4\eps^{-2})$. Furthermore, the quantum algorithm part only requires an additional ancilla qubit, two Hadamard gates, two $X$ gates, and two controlled gates $CG$ depending on the specific rotation gates $R_G$ used in the quantum circuit.

\subsection{Termination of the automated dimensional expressivity analysis and error estimates}\label{sec:automated-termination}

Since the proposed hybrid quantum-classical algorithm is an inductive procedure, we would like to say a few words about the base case, the termination condition, and error estimates.

Given the assumed parametric gates for the quantum circuit, the base case is trivial. $S_1$ is a $1\times1$-matrix and all diagonal elements of each $S_k$ are $\frac14$. Hence, $S_1$ is always invertible. 

To discuss the termination condition, it is advantageous to recognize that 
\begin{align}
  S_k=
  \begin{pmatrix}
    S_{k-1}&A_k\\
    A_k^*&\frac14
  \end{pmatrix}
	\label{eq:Sk_matrix}
\end{align}
holds with 
\begin{align}
  A_k:=J_{k-1}^*
  \begin{pmatrix}
    \Re\d_kC\\\Im\d_kC
  \end{pmatrix}
  =
  \begin{pmatrix}
    \frac{1}{4}\Re\langle\gamma_1,\gamma_k\rangle\\
    \vdots\\
    \frac{1}{4}\Re\langle\gamma_{k-1},\gamma_k\rangle
  \end{pmatrix}.
\end{align}
As such, $S_k$ is invertible if and only if $\frac14-A_k^*S_{k-1}^{-1}A_k$ is non-zero (using the Schur complement and the assumption that $S_{k-1}$ is invertible). Noting that $J_{k-1}$ has independent columns, we conclude that the Moore-Penrose pseudo-inverse\footnote{The Moore-Penrose pseudo-inverse of a matrix $A$ is defined as a matrix $A^+$ that satisfies $AA^+A=A$, $A^+AA^+=A^+$, $(AA^+)^*=AA^+$, and $(A^+A)^*=A^+A$.} $J_{k-1}^+$ is a left-inverse of $J_{k-1}$ and satisfies
\begin{align}
  J_{k-1}^+=S_{k-1}^{-1}J_{k-1}^*.
\end{align}
Furthermore, we obtain
\begin{align}
  A_k^*S_{k-1}^{-1}A_k=
  \begin{pmatrix}
    \Re\d_kC\\\Im\d_kC
  \end{pmatrix}^*
  J_{k-1}J_{k-1}^+
  \begin{pmatrix}
    \Re\d_kC\\\Im\d_kC
  \end{pmatrix}.
\end{align}
Since $J_{k-1}^+$ is only a left-inverse (unless in the case which will become the termination condition), the central $J_{k-1}J_{k-1}^+$ will not reduce to the identity. However, if $J_{k-1}^+$ is a right-inverse as well, then we observe
\begin{align}\label{eq:Schur-complement-Sk}
  \begin{split}
    \frac{1}{4}-A_k^*S_{k-1}^{-1}A_k=&\,\frac{1}{4}-
    \begin{pmatrix}
      \Re\d_kC\\\Im\d_kC
    \end{pmatrix}^*
    \begin{pmatrix}
      \Re\d_kC\\\Im\d_kC
    \end{pmatrix}\\
    =&\,\frac{1}{4}-\frac{1}{4}\langle\gamma_k,\gamma_k\rangle\\
    =&\,0.
  \end{split}
\end{align}

This termination condition should hold if the number of independent parameters coincides with the dimension $d$ of the physical state space, i.e., the sub-manifold of the quantum-device state space subject to all physical symmetries that the quantum circuit satisfies. In this case, the tangent space of the physical state space is $d$-dimensional and we can choose a representation $\mathcal{J}_k$ of $J_k$ such that $\mathcal{J}_k$ is a $d\times k$-matrix. In particular, $\mathcal{J}_d$ is a $d\times d$-matrix with independent columns. In other words, $\mathcal{J}_d$ is invertible and we conclude $\mathcal{J}_d^+=\mathcal{J}_d^{-1}$. \autoref{eq:Schur-complement-Sk} thus implies by induction that all following parameters will be rejected as dependent. A summary of the procedure in pseudo code is provided in \autoref{algo:hybrid_expressivity_algo_tobias}, where we generate both the independent and redundant parameters for pedagogical reasons.

\begin{algorithm}[ht]
  \DontPrintSemicolon
  \SetAlgoLined  
  \KwIn{Parametric circuit, $\theta_1,\dots, \theta_N$}
  \KwOut{Independent and redundant parameters}
  $I\leftarrow\{1\}$\;
  $R\leftarrow\emptyset$\;
  $S\leftarrow \frac{1}{4}$\;
    \For{$k\leftarrow 2$ \KwTo $N$}{
      \For{$l\leftarrow 1$ \KwTo $\#I$}{
        $l' \leftarrow I[l] $\;
        Measure $\frac{1}{4}\Re\langle\gamma_l',\gamma_k\rangle$ on the quantum device\;
        $A_k[l]\leftarrow \frac{1}{4}\Re\langle\gamma_l',\gamma_k\rangle$\;
      }
      $\tilde S\leftarrow \begin{pmatrix} S & A_k\\ A_k^* & \frac{1}{4}\end{pmatrix}$\;
      \eIf{$\tilde S$ is invertible}{
        $I\leftarrow I\cup\{k\}$\;
        $S\leftarrow \tilde S$\;
      }{
        $R\leftarrow R\cup\{k\}$ \;
      }
   }
    \KwRet{independent~parameters:~$\{\theta_i;\ i\in I\}$, redundant~parameters:~$\{\theta_r;\ r\in R\}$}\;        
   \caption{Hybrid quantum-classical expressivity analysis}
   \label{algo:hybrid_expressivity_algo_tobias}
\end{algorithm}

With respect to error estimates, we assume that all $\frac{1}{4}\Re\langle\gamma_m,\gamma_n\rangle$ are known up to an error of~$\eps$ (e.g., the corresponding multiple of the standard deviation for a given confidence level). The matrices $S_k$ are therefore estimated by matrices $S_k+\delta_k$, where each matrix element of $\delta_k$ is less than $\eps$ in absolute value (at least up to the chosen confidence level). If we choose to test $S_k$ for invertibility by computing its eigenvalues, then we can use the eigenvalue stability theorem
\begin{align}
  \abs{\mathrm{error}(\lambda)}\le\norm{\delta_k}_F\le k\eps\le N\eps
\end{align}
for all eigenvalues $\lambda$ of $S_k$, where 
\begin{align}
  \norm{T}_F:=\sqrt{\sum_{i,j}\abs{T_{i,j}}^2}
\end{align}
denotes the Frobenius norm\footnote{The Frobenius norm is also known as the Hilbert-Schmidt norm and sometimes called Euclidean norm.} of the matrix $T$. Hence, if the smallest eigenvalues $\hat\lambda_{\min}$ of $S_k+\delta_k$ is compatible with zero, i.e., $\hat\lambda_{\min}\le k\eps$ (note that $S_k=J_k^*J_k$ is non-negative), then we would reject the currently tested parameter as being dependent on the previous set of parameters.

With this estimate $\hat\lambda_{\min}\le k\eps$, we could in principle falsely identify a parameter as dependent when it is indeed independent. However, as increasing the number of measurements for the evaluation of $\frac{1}{4}\Re\langle\gamma_m,\gamma_n\rangle$ reduces $\eps$, we can identify independent parameters once sufficiently many quantum device calls have been used to ensure that $\hat\lambda_{\min}>k\eps$. Of course, this is still subject to the chosen confidence level but it shows that the probability of falsely identifying a dependent parameter as independent can be made arbitrarily small. Conversely, a dependent parameter should always have a minimal eigenvalue estimate $\hat\lambda_{\min}\le k\eps$. 

Choosing other tests for invertibility, such as $\frac14-A_k^*S_{k-1}^{-1}A_k>0$,\footnote{Note that $J_{k-1}J_{k-1}^+$ satisfies $(J_{k-1}J_{k-1}^+)^*=J_{k-1}J_{k-1}^+$ and $(J_{k-1}J_{k-1}^+)^2=J_{k-1}J_{k-1}^+J_{k-1}J_{k-1}^+=J_{k-1}J_{k-1}^+$. Hence, $J_{k-1}J_{k-1}^+$ is an orthoprojector onto the column space of $J_{k-1}$. In particular, we obtain $0\le J_{k-1}J_{k-1}^+\le1$ and, setting $v=\begin{pmatrix}\Re\d_kC\\\Im\d_kC\end{pmatrix}$, we conclude $0\le A_k^*S_{k-1}^{-1}A_k=v^*J_{k-1}J_{k-1}^+v\le\norm{v}_2^2=1/4.$} also allow for error estimates of the form $ck\eps$ where $c$ is an $\eps$-dependent constant larger than $c_0:=2\norm{S_{k-1}^{-1}}\norm{A_k}_2+\norm{S_{k-1}^{-1}}^2\norm{A_k}_2^2$ and $c_0$ holds as a bound in the $\eps\to0$ limit. 
Choosing $\det(S_k)>0$ as a test is not recommended since $0\le\det(S_k)\le 4^{-k}$ holds. This follows by induction using $\det(S_1)=\frac14$, $\det(S_k)=\det(S_{k-1})\det\left(\frac14-A_k^*S_{k-1}^{-1}A_k\right)$, and $0\le \frac14-A_k^*S_{k-1}^{-1}A_k\le\frac{1}{4}$. In any case, we see that there are numerous ways to identify independent parameters.

Since we compute $S_k$ using a quantum device, any test for parameter dependence/independence needs to be resistant to perturbations introduced by the quantum part of the algorithm. The parameter \textit{independence} test satisfies this perturbation resistance because $\hat\lambda_{\min}>k\eps$ will be satisfied if we can sufficiently reduce the noise level~$\eps$. The parameter \textit{dependence} test, on the other hand, needs to test $\det(S_k)=0$ or, equivalently, test that the final column vector in $J_k$ is contained in the subspace spanned by the first $k-1$ columns of $J_k$. Both of these tests ask whether some mathematical object is an element of a zero-measure set. Once the perturbations from the quantum device are introduced, the probability of positively identifying a dependent parameter as being dependent is therefore zero. In other words, without perturbation resistant sufficient conditions for parameter dependence, we need to introduce threshold conditions and either accept that we may erroneously consider a dependent parameter to be independent or vice versa.

Erring on the side of caution, in line with any chosen confidence level (interpreting $\eps$ as a suitable multiple of the standard deviation of the quantum algorithm part), we therefore might exclude an independent parameter by erroneously labeling it as dependent. This has the advantage that we do not consider this parameter to increase expressivity, i.e., if it does increase expressivity then the circuit is indeed more expressive than we have concluded. The alternative would be to erroneously allow dependent parameters to be declared independent. This would lead to the problem of constructing a supposedly maximally expressive circuit, but some states cannot be generated because we falsely counted the dimension of the circuit manifold. Of course, while this choice ensures maximal expressivity, it may lead to suboptimal circuit reductions.

\subsection{Implementation on IBMQ}\label{sec:hardware-implementation}

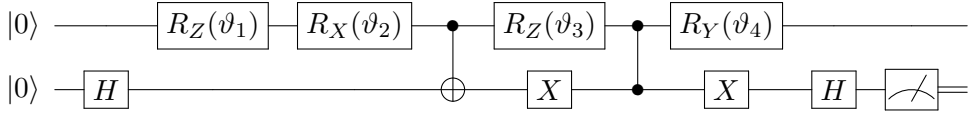
\begin{figure*}
  \begin{align*}
    \Qcircuit @C=1em @R=.7em {
      \lstick{\ket{0}} & \qw      & \gate{R_Z(\theta_1)} & \gate{R_X(\theta_2)} & \ctrl{1} & \gate{R_Z(\theta_3)} & \ctrl{1} & \gate{R_Y(\theta_4)} & \qw      & \qw    & \qw \\
      \lstick{\ket{0}} & \gate{H} & \qw                  & \qw                  & \targ    & \gate{X}             & \ctrl{0} & \gate{X}             & \gate{H} & \meter & \cw \\
      }
  \end{align*}
  \caption{Circuit for obtaining $\Re\langle\gamma_2,\gamma_3\rangle$ on quantum hardware for the ansatz in \autoref{eq:single_qubit_experiment}. The ancillary qubit is initially put into superposition and acts as a control for the additional CNOT and $CZ$ gates. After applying again a Hadamard on the ancilla, a final measurement reveals the probability for the qubit to be in state $\ket{0}$, which is related to $\Re\langle\gamma_2,\gamma_3\rangle$ according to \autoref{eq:real_part_overlap}.}\label{fig:Ak_circuit_single_qubit}
\end{figure*}
\begin{figure}[t]
	\centering
  \includegraphics[width=0.48\textwidth]{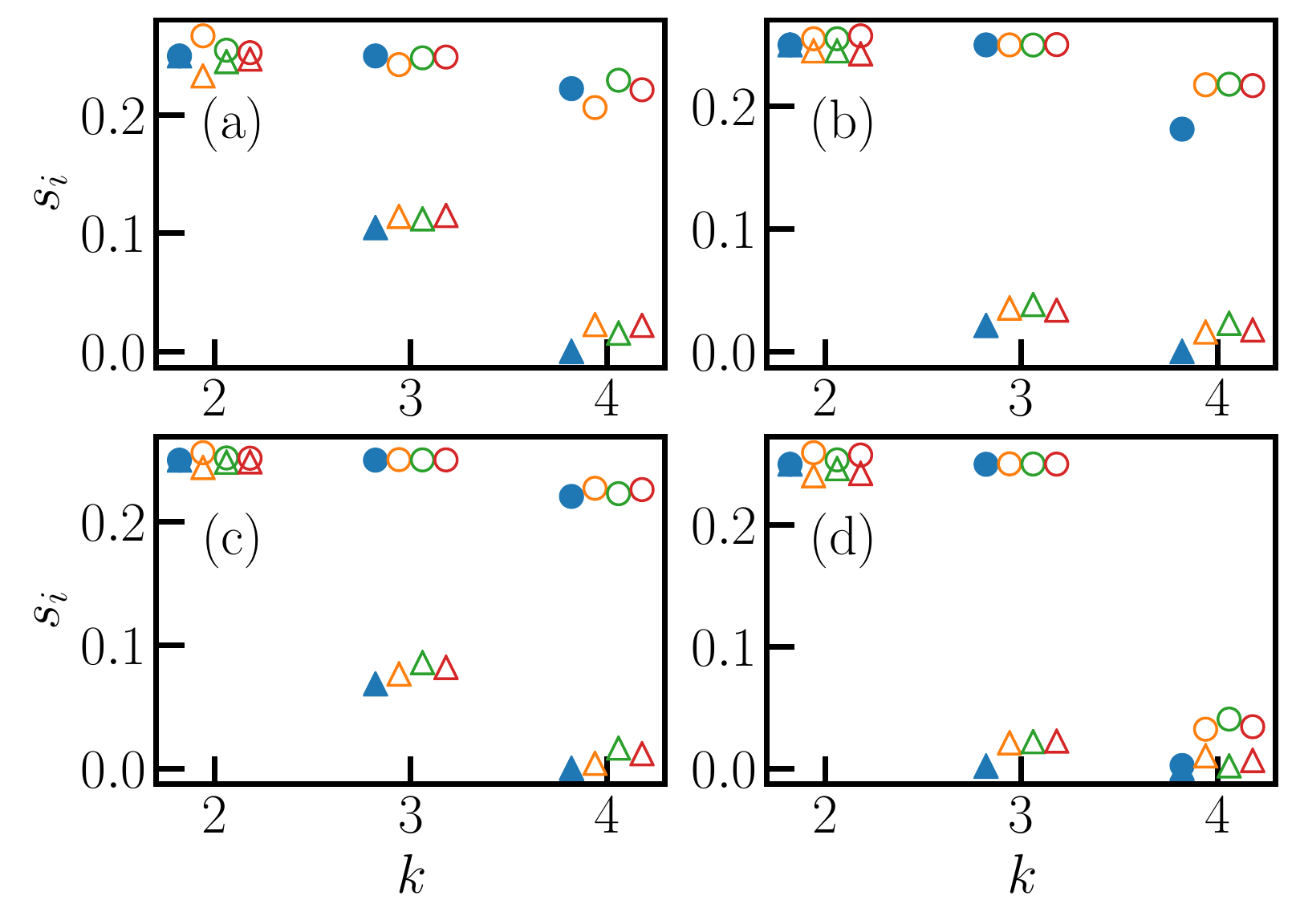}
	\caption{Smallest (triangles) and second smallest (dots) eigenvalues of the matrices $S_k$, $k\geq 2$, for the circuit in \autoref{eq:single_qubit_experiment}. The different panels each correspond to a different, randomly drawn parameter set $\theta_1,\dots,\theta_4$. The filled markers indicate the exact solution, the open markers the results obtained from ibmq\_vigo with 1000 measurements (orange markers), 4000 measurements (green markers) and 8000 measurements (red markers). The data points for each $k$ are slightly horizontally shifted for better visibility.}
	\label{fig:single_qubit_Sk_results_ibmq_vigo}
\end{figure}

In order to experimentally test our approach, we implement the algorithm on IBM's quantum hardware using QISKIT~\cite{Abraham2019}. As described above, we can measure the entries of the vectors $A_k$ on a quantum chip, which, starting from $S_1=1/4$, allows us to recursively compute the matrices $S_k$ using \autoref{eq:Sk_matrix}. Subsequently, we can numerically determine the spectra of these matrices which allows us to identify superfluous parameters. For comparison, we also construct the exact $S_k$ matrices in order to benchmark our hardware results.

As an example, we consider the single-qubit circuit
\begin{align}
	\begin{aligned}
		C(\theta_1, &\theta_2, \theta_3, \theta_4) =\\ 
		&R_Y(\theta_4)R_Z(\theta_3)R_X(\theta_2)R_Z(\theta_1)\ket{0}.
	\end{aligned}
	\label{eq:single_qubit_experiment}
\end{align}
Inspecting \autoref{eq:single_qubit_experiment}, it is apparent that the first $R_Z$ gate can generate arbitrary phases, while the following two rotations $R_X$ and $R_Z$ allow us to map the initial state $\ket{0}$ to any point on the Bloch sphere. As a result, the last $R_Y$ gate is superfluous and we expect the matrix $S_4$ to have a single vanishing eigenvalue, while the spectrum for all other $S_k$, $k\leq 3$, should consist of nonzero values. To check if this is indeed the case we choose various random parameter sets $(\theta_1, \theta_2, \theta_3, \theta_4)$, where each angle is drawn uniformly from the interval $[0,2\pi)$, and evaluate the spectra of the $S_k$ matrices (note that $S_1$ is always $1/4$ and thus does not need to be measured). To construct the matrices $S_k$, we measure the vector $A_k$ on the quantum device as outlined in \autoref{sec:hybrid-algorithm}. An example for a circuit allowing us to obtain one of the entries of $A_k$ for our ansatz in \autoref{eq:single_qubit_experiment} is shown in \autoref{fig:Ak_circuit_single_qubit}.

Our data for the low-lying spectrum obtained for different parameter sets are shown in \autoref{fig:single_qubit_Sk_results_ibmq_vigo} and \autoref{tab:single_qubit_Sk_results_ibmq_vigo}.
Looking at the exact solution in \autoref{fig:single_qubit_Sk_results_ibmq_vigo} and \autoref{tab:single_qubit_Sk_results_ibmq_vigo}, we see indeed that $S_4$ has a single vanishing eigenvalue for all the parameter sets we study, in agreement with the theoretical consideration. Focusing on the cases in \autoref{fig:single_qubit_Sk_results_ibmq_vigo}(a) - \autoref{fig:single_qubit_Sk_results_ibmq_vigo}(c), we observe that the smallest eigenvalue of $S_3$ is clearly distinct from zero, thus allowing us to reliably identify the parameter $\theta_4$ as superfluous. \autoref{fig:single_qubit_Sk_results_ibmq_vigo}(d) shows the results for a parameter set which is close to a singular point and at first glance both $S_3$ and $S_4$ seem to have a zero eigenvalue. A closer examination of the numerical values in \autoref{tab:single_qubit_Sk_results_ibmq_vigo} reveals that for the exact solution the smallest eigenvalue of $S_3$ is on the order of $10^{-3}$, whereas $S_4$ has a vanishing eigenvalue. Thus, we can still confidently identify the redundant gate in the circuit. By construction, $S_k$ has a single vanishing eigenvalue if and only if $\vartheta_k$ is first redundant parameter in the circuit. Our data also show that, apart from the parameter set close to a singular one, there is a clear separation between the smallest and the second smallest eigenvalue for $S_k$, $k\geq 3$. Hence, even if one can only resolve the spectra of the $S_k$ matrices to a finite precision (e.g. due to a finite number of measurements or noise on a real device), we expect the signal is not immediately lost.

Looking at our hardware results obtained on ibmq\_vigo, we see that this is indeed the case and, in general, we observe good agreement between the data from the quantum chip and the exact solution. The data from the quantum device do not show a strong dependence on the number of measurements, and already with a modest number of 1000 measurements, we are able to determine the lowest eigenvalues of the $S_k$ matrices with satisfactory precision. Due to the noise in the quantum device, some of the eigenvalues are slightly shifted compared to the exact result. 

In particular, the spectrum of $S_4$ does in general not contain an eigenvalue that is exactly zero, as \autoref{tab:single_qubit_Sk_results_ibmq_vigo} shows. Looking at \autoref{fig:single_qubit_Sk_results_ibmq_vigo}(b) and \autoref{fig:single_qubit_Sk_results_ibmq_vigo}(d), we see that these shifts in the spectrum might affect the determination of the redundant parameter. In  \autoref{fig:single_qubit_Sk_results_ibmq_vigo}(d), the shifts due to noise lead to a clear separation between the smallest eigenvalues of $S_3$ and $S_4$, although the parameter set chosen is close to a singular one. On the contrary, in \autoref{fig:single_qubit_Sk_results_ibmq_vigo}(b), the effects of noise cause the smallest eigenvalues of $S_3$ and $S_4$ to slightly deviate from zero and become more similar (see \autoref{tab:single_qubit_Sk_results_ibmq_vigo}). 

As a result, noise in current quantum devices might lead to misidentifying the superfluous parameter in certain cases. Nevertheless, considering all our hardware results for the different parameter sets, the smallest eigenvalue of $S_4$ is consistently the closest one to zero. Thus, despite the noise in ibmq\_vigo, we can confidently identity $\theta_4$ as superfluous parameter. We also observe a similar performance for different IBM quantum devices (see \aref{app:hardware_implementation} for details).

Although we focus on a simple single-qubit example here for illustration, our algorithm can readily be applied to more complex circuits, as for example QISKIT's \verb|EfficientSU2| circuit, which we discussed previously in \autoref{sec:EfficientSU2} (see \aref{app:hardware_implementation} for details).

\begin{table}[t]
  \centering
  \begin{tabular}{|l||c|c|c|c|}
  \hline
   & (a) & (b) & (c) & (d)  \\ \hline \hline
  $s_2$ exact & 0.250 & 0.250 & 0.250 & 0.250\\ \hline
  $s_2$ ibmq & 0.247 & 0.243 & 0.248 & 0.242\\ \hline \hline
  $s_3$ exact & 0.105 & 0.021 & 0.068 & 0.002\\ \hline
  $s_3$ ibmq & 0.115 & 0.034 & 0.082 & 0.022\\ \hline \hline
  $s_4$ exact & 0 & 0 & 0 & 0\\ \hline
  $s_4$ ibmq & 0.022 & 0.018 & 0.012 & 0.007\\ \hline
  \end{tabular}  
  \caption{Numerical values for the smallest eigenvalues~$s_k$ of the $S_k$ matrices for the exact case and the hardware results obtained with 8000 shots. The different columns correspond the different parameter sets shown in the panels of \autoref{fig:single_qubit_Sk_results_ibmq_vigo}.}
  \label{tab:single_qubit_Sk_results_ibmq_vigo}
\end{table}

\section{Conclusion}\label{sec:conclusion}
In this paper, we proposed a dimensional expressivity analysis for the study and design of parametric quantum circuits. This is of particular interest if the quantum circuit is intended to obey certain physical symmetries and to avoid others. The dimensional expressivity analysis relies on the comparison between the tangent spaces of the physical state space and the manifold of reachable states by the quantum circuit. Fundamentally, a quantum circuit is maximally expressive if and only if the tangent spaces coincide, as then the circuit manifold coincides with the physical state space. Furthermore, identification of a basis in the tangent space allows for the removal of redundant parameters in the quantum circuit. This removal of redundant parameters can also be extended to remove unwanted symmetries from the states generated by the quantum circuit. 

We have applied the dimensional expressivity analysis (developed in \autoref{sec:parameter_reduction}) to various quantum circuits. As a simple introductory example, we have removed all redundant parameters from QISKIT's \verb|EfficientSU2(3, reps=N)| circuit in \autoref{sec:EfficientSU2}. The analysis therefore provided a maximally expressive $3$-qubit quantum circuit that is free from redundant parameters. Additionally, we have optimized QISKIT's \verb|EfficientSU2(3, reps=N)| circuit by removing its ability to generate an arbitrary phase. 

Moreover, we have developed a hybrid-classical algorithm, which allows us to efficiently perform the dimensional expressivity analysis using quantum hardware. We have implemented the algorithm on IBM's quantum hardware and demonstrated its performance for the single (\autoref{sec:hardware-implementation}) and two-qubit (\aref{sec:hardware-implementation-2Q}) cases. Although we have not applied any error mitigation~\cite{Temme2017,Kandala2017,Endo2018,YeterAydeniz2020,Funcke2020}, we can identify the superfluous parameters reasonably well in both cases we study.

Furthermore, we provided an example for dimensional expressivity analysis in the presence of physical symmetries. Using translationally invariant Hamiltonians as an illustration, we analyzed the structure of the physical state space in \autoref{sec:translation_invariance} and \aref{app:eigenspaces_path_connected}. The specific details of this analysis serve as a guide to similar analyses for general symmetries generated from other spectral invariants.

In addition to the knowledge of maximal expressivity of a quantum circuit, or deficiency thereof, the analysis of physical sectors provides valuable information that can guide the custom design of maximally expressive quantum circuits. We have provided such an example in \autoref{sec:4Q_circuit_design}, where we analyzed a non-performing $4$-qubit translationally invariant circuit, understood why the VQS failed to find the ground state, and used the information to design a maximally expressive, translationally invariant $4$-qubit circuit. We then showed that this custom circuit overcame the difficulties encountered by the initial ansatz.

In this paper, we mainly used dimensional expressivity analysis with regard to VQS. 
The method developed is however completely general and can be applied to any parametric quantum circuit. It allows us to (i) decide whether or not a quantum circuit is maximally expressive, (ii) remove any redundant parameters from a given quantum circuit improving circuit efficiency, (iii) obtain a measure of deficiency in the number of parameters necessary to obtain a maximally expressive quantum circuit, (iv) design custom gate sets satisfying the physical symmetries of the physical model, and (v) remove any unwanted symmetries that a candidate quantum circuit may contain.

\begin{acknowledgments}
Research at Perimeter Institute is supported in part by the Government of Canada through the Department of Innovation, Science and Industry Canada and by the Province of Ontario through the Ministry of Colleges and Universities. S.K.\ acknowledges financial support from the Cyprus Research and Innovation Foundation under project "Future-proofing Scientific Applications for the Supercomputers of Tomorrow (FAST)", contract no.\ COMPLEMENTARY/0916/0048. P.S. thanks the Helmholtz Einstein International Berlin Research School in Data Science (HEIBRiDS) for funding. We acknowledge the use of IBM Quantum services for this work. The views expressed are those of the authors, and do not reflect the official policy or position of IBM or the IBM Quantum team.
\end{acknowledgments}

\appendix

\section{Path connectedness of momentum sectors}\label{app:eigenspaces_path_connected}

The incorporation of problem specific symmetries and thus the reduction of the full state space (unit sphere in a complex Hilbert space) to physical sectors generated by the symmetry (here the intersection of the eigenspaces of the translation operator with the unit sphere) may disconnect the physical sectors into multiple connected components. In other words, there may be multiple topological sectors in each physical sector. For example, if there is a ``hidden'' $\zn_2$ symmetry, then a given starting position may prevent the optimizer from finding the global optimum if the quantum circuit fixes parity. In \aref{sec:path_connected_proof}, we will show that this is not the case for the momentum sectors. In particular, we will show that each momentum sector is path connected and that there is a path connecting any two states in any given momentum sector. In fact, the proof makes no reference to the translation operator per se. As such, it holds for all non-trivial spectrally invariant sectors, that is, physical sectors that are intersections of eigenspaces (with dimension $\ge2$) of a given operator and the unit sphere in the Hilbert space. Furthermore, assuming the parameter space is path connected and the quantum circuit is surjective (with respect to the physical sector), then any state in the physical sector can be connected to an initial state generated by an initial parameter set using a path in parameter space. This will be discussed in \aref{sec:path_construction}.

These results are not only important as they guarantee the ability of a maximally expressive quantum circuit to reach any state in the physical sector. They are particularly interesting if we suspect the quantum circuit to contain a ``hidden'' discrete symmetry. While a hidden continuous symmetry reduces the circuit manifold dimension, thus preventing the construction of a maximally expressive circuit, a discrete symmetry is more likely to result in the physical sector splitting into multiple connected components. If we are confronted with a situation in which a maximally expressive quantum circuit cannot generate a path connecting two arbitrary states in the physical sector, then this may be due to an unintentional discrete symmetry all states generated by the circuit obey.

Unfortunately, this does not always have to be caused by a symmetry. The set of singular points, in which there are jumps from lower-semicontinuity, can theoretically form a lower dimensional manifold which splits the physical sector into different regions. A gradient flow would not be able to pass through such a lower dimensional manifold of singular points, i.e., any gradient based solver is unlikely to cross between such regions as well (even if each step is subject to small perturbations unless the solver is trapped on the manifold of singular points).

In either case, failure of path connectedness of a maximally expressive quantum circuit mapping into a spectrally invariant physical sector indicates a fundamental necessity for an in-depth analysis of the quantum circuit in relation to the physical model.

\subsection{Proof of path connectedness}\label{sec:path_connected_proof}
To prove the path connectedness  claims, let $T$ be an operator, $\lambda$ an eigenvalue of $T$, $\ket\phi$ and $\ket\psi$ be two normalized eigenvectors of $T$ with respect to $\lambda$. First, we need to construct a path $\gamma$ such that $\gamma(0)=\ket\phi$ and $\gamma(1)=\ket\psi$ to show that the physical sector is path connected. Since the dimension of the eigenspace is $\ge2$, we can assume without loss of generality that $\ket\phi$ and $\ket\psi$ are linearly independent (otherwise, we can find a linearly independent~$\ket\chi$ and connect $\ket\chi$ to both $\ket\phi$ and $\ket\psi$). To construct a path from $\ket\phi$ to $\ket\psi$, we need to orthonormalize, that is, we need to find $\ket\xi=\alpha\ket\phi+\beta\ket\psi$ with 
\begin{align}
  \begin{split}
    1=\,&\langle\xi|\xi\rangle \\
    =\,&\abs\alpha^2 + \alpha^*\beta\langle\phi|\psi\rangle+\alpha\beta^*\langle\psi|\phi\rangle+\abs\beta^*
  \end{split}
\end{align}
and
\begin{align}
  0=\,&\langle\phi|\xi\rangle=\alpha+\beta\langle\phi,\psi\rangle.
\end{align}
Inserting $\alpha=-\beta\langle\phi,\psi\rangle$ into the first equation yields
\begin{align}
  1=\,&\abs\beta^2\l(1-\abs{\langle\phi,\psi\rangle}^2\r).
\end{align}
Setting $\alpha:=-\frac{\langle\phi,\psi\rangle}{\sqrt{1-\abs{\langle\phi,\psi\rangle}^2}}$ and $\beta:=\frac{1}{\sqrt{1-\abs{\langle\phi,\psi\rangle}^2}}$ provides a suitable choice of $\ket\xi$. The function
\begin{align}
  \begin{split}
    \gamma_0(t):=\,&\cos(t) \ket\phi+\sin(t)\ket\xi\\
    =\,&\l(\cos(t)-\frac{\langle\phi,\psi\rangle \sin(t)}{\sqrt{1-\abs{\langle\phi,\psi\rangle}^2}}\r)\ket\phi\\
    &+\frac{\sin(t)}{\sqrt{1-\abs{\langle\phi,\psi\rangle}^2}}\ket\psi
  \end{split}
\end{align}
thus satisfies $\gamma_0(0)=\ket\phi$ and $\gamma_0(t_\psi)=\ket\psi$ for $t_\psi:=\arcsin\l(\sqrt{1-\abs{\langle\phi,\psi\rangle}^2}\r)$. We therefore observe that $\gamma(t):=\gamma_0(t_\psi t)$ is a path connecting $\ket\phi$ and $\ket\psi$, every $\gamma(t)$ has norm $1$, and every $\gamma(t)$ is an eigenvector of $T$ with respect to $\lambda$. This proves that the physical sectors are path connected.

\subsection{Paths in parameter space}\label{sec:path_construction}
Let us now assume that the quantum circuit~$C$ maps the parameter space into the physical sector surjectively and that the parameter space is path connected. Given a state $\ket\psi$, surjectivity of $C$ ensures the existence of a parameter set $\theta$ such that $\ket\psi=C(\theta)$. Fixing an initial parameter set $\theta_0$, we can find a path $\gamma_0$ in the parameter space connecting $\theta_0$ to $\theta$. By continuity of $C$, we therefore obtain a path $\gamma:=C\circ\gamma_0$ in the physical sector connecting the initial state $C(\theta_0)$ to $\ket\psi$.

Although this guarantees the existence of a path in parameter space and therefore the theoretical ability for the circuit $C$ to find the optimum using a local optimizer, if we try to construct this path using information obtained in the physical sector (as we do in variational quantum simulations), we can see the problem of singular points. To illustrate this observation, we will first describe the construction of such a path if there are no singular points on the path.

Starting with a path $\gamma$ in the physical sector, we would like to find a path $\gamma_0$ in parameter space such that $\gamma=C\circ\gamma_0$; i.e., we need to invert~$C$. This is possible if we restrict the circuit to its minimal independent set of parameters at each point in parameter space. If there are no singular points on the path, then the perturbation estimates of \autoref{sec:semicontinuity} provide a radius of validity for each restriction at points on $\gamma$. By compactness of $[0,1]$, the image of $\gamma$ can be split into finitely many pieces. On each of these pieces, we can use one restriction $\check C$ of $C$ to obtain $\gamma_0=\check C^{-1}\circ\gamma$.

Some care has to be taken at the point switching from one restriction to another because the parameters held constant may not coincide. But since the parameters held constant are dependent on the independent parameters, they can be varied to match up without changing the state at which the switch is performed. Once the parameters held constant coincide, the remaining parameters can be made to match up. This inserts a ``detour'' into the path $\gamma$ but since we assume that $C$ does not map outside the physical sector, we still have a path connecting the two initial states. Ensuring that the first restriction corresponds to the initial parameter set and the final restriction to the final parameter set, we have found a path in parameter space from $\theta_0$ to $\theta$ that connects $\ket\phi$ to $\ket\psi$.

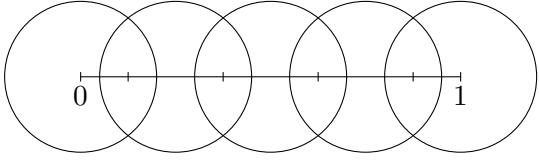
\begin{figure}[t]
  \begin{center}
    \begin{tikzpicture}[scale=2.5]
      \draw (-1,0) -- (1,0);
      \draw (-1,-.025) -- (-1,.025);
      \draw (1,-.025) -- (1,.025);
      \foreach \i in {0,...,4}{
        \draw (-1+.5*\i,0) circle (.4);
      }
      \foreach \j in {0,...,3}{
        \draw (-1+.5*\j+.25,-.025) -- (-1+.5*\j+.25,.025);
      }
      \node[below] (note) at (-1,0) {$0$};
      \node[below] (note) at (1,0) {$1$};
    \end{tikzpicture}
  \end{center}
  \caption{\label{fig:overlapping_radii_of_validity_regular} Visualization of a path in physical space with radii of validity for restricted circuits. No singular points are present. Thus there are finitely many overlapping areas of validity (circles). Points to swap to the next restriction are in the interior of the intersection of two areas of validity.}
\end{figure}

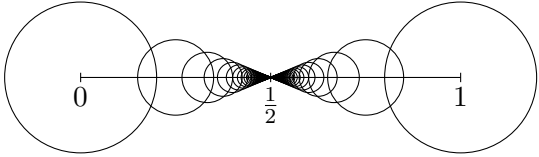
\begin{figure}[t]
  \begin{center}
    \begin{tikzpicture}[scale=2.5]
      \draw (-1,0) -- (1,0);
      \draw (-1,-.025) -- (-1,.025);
      \draw (0,-.025) -- (0,.025);
      \draw (1,-.025) -- (1,.025);
      \foreach \i in {1,...,200}{
        \draw (-1/\i,0) circle (.4/\i);
        \draw (1/\i,0) circle (.4/\i);
      }
      \node[below] (note) at (-1,0) {$0$};
      \node[below] (note) at (1,0) {$1$};
      \node[below] (note) at (0,0) {$\frac12$};
    \end{tikzpicture}
  \end{center}
  \caption{\label{fig:overlapping_radii_of_validity_singular} Visualization of a path in physical space with radii of validity for restricted circuits for construction. The point $\frac12$ is singular. The radii of validity therefore converge to zero on either side. This separates the interval into two pieces without overlapping areas of validity (circles). }
\end{figure}

The construction is therefore intimately relies on overlapping radii of validity for the restricted circuits in \autoref{fig:overlapping_radii_of_validity_regular}. At a singular point, these radii converge to zero. The situation for a singular point at $\frac12$ is visualized in \autoref{fig:overlapping_radii_of_validity_singular} instead. In the case without singular points, the points at which restricted circuits are changed lie within the area of validity of both restricted circuits. This allows for local information to be utilized. At the singular point, however, only pointwise information is available if the path is not contained in the lower dimensional manifold generated by the restricted circuit at the singular point.

Luckily, having a singular point on the path is not always a calamity. Although a gradient based optimizer will be particularly vulnerable, it is not guaranteed to fail. Due to discrete step sizes, there is a chance that an update will step from one side of the singularity to the other. If this is the case, it is unlikely to be any more noticeable than a slowing down of the optimizer near the singularity. However, if we consider a gradient descent, in which the update step is a scaled multiple of the gradient, a scaling too small would ensure that this stepping over the singularity will not happen. A possible result of this would be that the gradient descent follows the manifold of singular points, slightly off-set on one side, never crossing to the singularities.

Similarly, if the path is orthogonal to the tangent space of the restricted circuit at the singular point, then too small a gradient scaling would result in the gradient descent approximating the singular point and terminating because the norm of the gradient is below the termination condition. A gradient based optimizer would therefore terminate ``successfully'' in this case and proclaim the singular point to be the optimum. Having no further insight into the expressivity of the quantum circuit, this situation would be indistinguishable from the optimizer terminating with a false solution due to a local optimum.

Although counter-intuitive, this suggests that increasing the scaling for the parameter update of a local optimizer near an apparent solution can be useful. If it is truly a local optimum and the scaling ``overshoots'' the update, then we expect to see updates bouncing around the local optimum. If, on the other hand, the slowing down is due to a singular point nearby, then ``overshooting'' means the artificial optimum is passed and the optimizer stops behaving like it would close to terminating.

The radius of validity estimate $R\ge\min\{R_0,\delta^{-1}\abs{\det(C_0(\theta_0))}\}$ derived in \autoref{sec:semicontinuity} can also be used to identify singular points nearby. Approaching a singular point implies convergence of $R$ to $0$. Being in a neighborhood of a non-singular local optimum should not show $R\to0$. In fact, $R$ can be estimated using $\theta\mapsto\abs{\det(C_0(\theta))}$ or $\theta\mapsto\frac{\abs{\det(C_0(\theta))}}{\norm{(\det\circ C_0)'(\theta)}_{\ell_2}}$. Both of these expressions converge to $0$ in a neighborhood of a singular point as well ($\abs{\det(C_0(\theta))}=0$ is characteristic for loss of independent parameters due to lower-semicontinuity, and $\frac{\abs{\det(C_0(\theta))}}{\norm{(\det\circ C_0)'(\theta)}_{\ell_2}}$ is a first order approximation of $R$ at $\theta$). 

Hence, while path connectivity of spectrally invariant sectors is a very useful property ensuring that a VQS with local update routine is theoretically capable of resolving the solution state, provided the quantum circuit is maximally expressive, constructing this path from an initial parameter set to a parameter set corresponding to the solution can be more difficult. In particular, singular points in the maximal expressivity of a quantum circuit can become problematic if this behavior is not monitored.

To monitor proximity to singular points and thereby induced failure of the VQS, we can estimate the radius of validity for the current reduced quantum circuit. In general, if we are have a quantum circuit whose reductions have singular points, we are therefore looking for multiple reduced circuits $\check C_j$. Each of the $\check C_j$ should have an open and path connected domain $P_j$ and each point $\theta$ in parameter space should be contained in at least one of the $P_j$. Since the radius of validity at each point $\theta$ in $P_j$ is bounded from below by the distance $\dist(\theta,\d P_j)$ of $\theta$ to the boundary of $P_j$ and this distance is a lower bound for the distance to any singular points of $\check C_j$, practical methods of avoiding singular points exist if this path connectivity monitoring is included in the VQS. For example, if only finitely many $\check C_j$ are used, then one could always choose the circuit $\check C_j$ for which the current point $\theta$ is contained in $P_j$ and $\dist(\theta,\d P_j)$ is maximal. 

In this way, monitoring the path of the VQS as well as distances to singular points can circumvent many problems arising from circuit induced local minima.

\section{Details of the implementation of the hardware efficient algorithm\label{app:hardware_implementation}}
In the main text, we provided a proof of principle demonstration for a single-qubit case using ibmq\_vigo. Here we give further information about the simulation parameters. In addition, we also assess the performance of various IBM quantum hardware in greater detail and also apply our hardware efficient algorithm to QISKIT's \verb|EfficientSU2| circuit for 2 qubits.

In general, the topology of the IBM devices does not offer all-to-all connectivity of the qubits. The topology of the devices we use is given by a linear chain of 3 qubits of which the middle qubit is connected to another linear chain of 2 qubits, except for ibmq\_santiago which is a linear chain of 5 qubits~\cite{IBMQuantum2020}. Thus, in a first step we use the transpiler of QISKIT to translate the circuits into a format compatible with the topology of the device. For all the results in the main text and this section we use the highest optimization level to achieve an optimal layout.

\subsection{Single-qubit example, various quantum hardware}
\begin{figure}[hb!]
	\centering
  \includegraphics[width=0.48\textwidth]{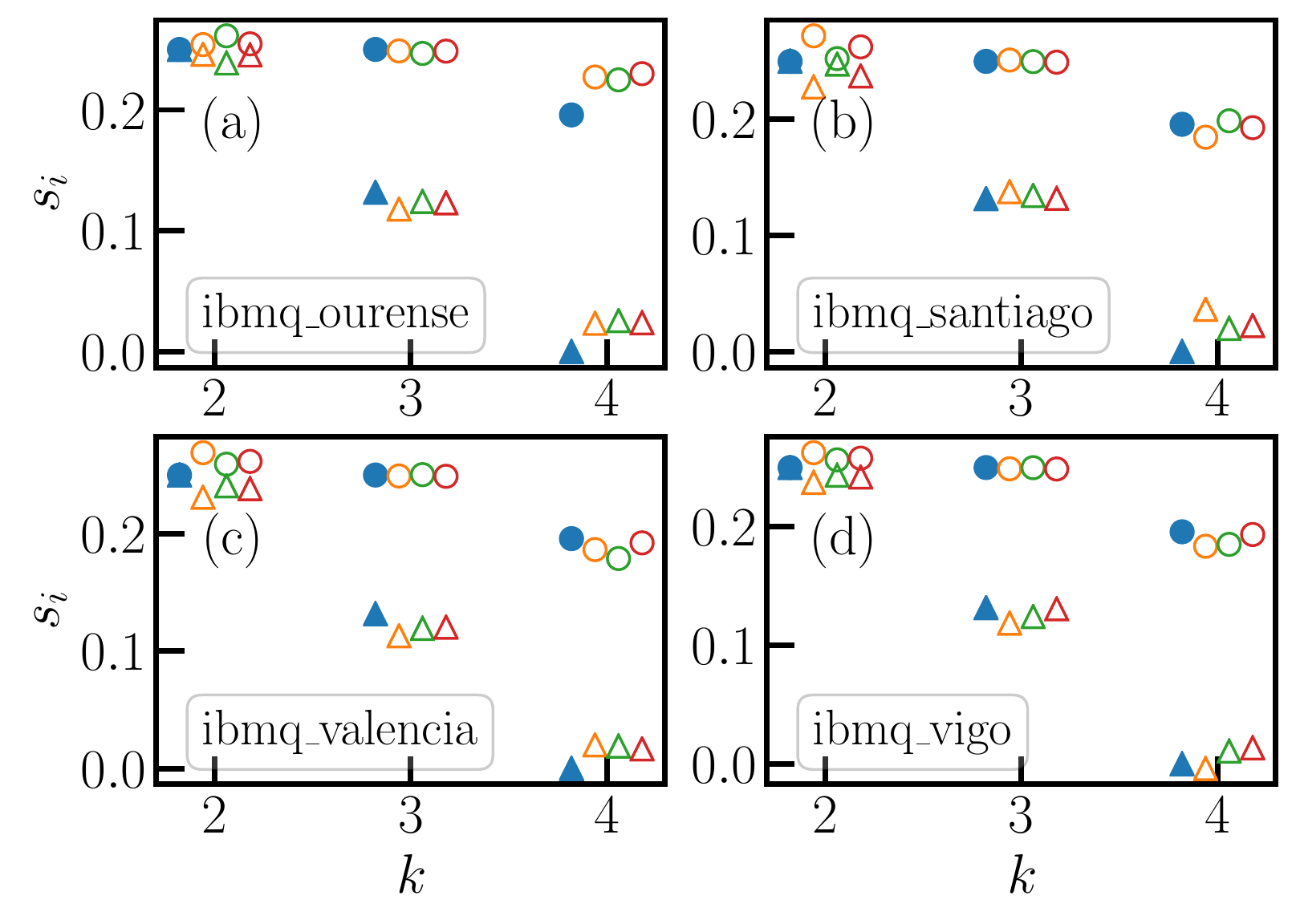}
	\caption{Smallest (triangles) and second smallest (dots) eigenvalues of the matrices $S_k$, $k\geq 2$, for the circuit in \autoref{eq:single_qubit_experiment}. The different panels correspond to a randomly drawn parameter set evaluated on the different hardware backends ibmq\_ourense (a), ibmq\_santiago (b), ibmq\_valencia (c), and ibmq\_vigo (d). The filled markers indicate the exact solution, the open markers the results obtained from the quantum hardware with 1000 measurements (orange markers), 4000 measurements (green markers) and 8000 measurements (red markers). The data points for each $k$ are slightly horizontally shifted for better visibility.}
	\label{fig:single_qubit_Sk_results}
\end{figure}

In \autoref{sec:hardware-implementation} in the main text, we focused on our results for ibmq\_vigo. To assess the performance of different quantum hardware, we repeat the single-qubit experiment using the circuit in \autoref{eq:single_qubit_experiment} for randomly drawn parameter sets on different chips. Our results are shown in \autoref{fig:single_qubit_Sk_results} and \autoref{tab:single_qubit_Sk_results}.

Comparing the results obtained from the different hardware backends, we observe only minor

\begin{table}[hb!]
  \centering
  \begin{tabular}{|l||c|c|c|c|}
  \hline
    & (a) & (b) & (c) & (d)  \\ \hline \hline
    $s_2$ exact & 0.250 & 0.250 & 0.250 & 0.250\\ \hline
    $s_2$ ibmq & 0.245 & 0.238 & 0.238 & 0.242\\ \hline \hline
    $s_3$ exact & 0.132 & 0.132 & 0.132 & 0.132\\ \hline
    $s_3$ ibmq & 0.123 & 0.132 & 0.120 & 0.131\\ \hline \hline
    $s_4$ exact & 0 & 0 & 0 & 0\\ \hline
    $s_4$ ibmq & 0.024 & 0.022 & 0.017 & 0.014\\ \hline
  \end{tabular}  
  \caption{Numerical values for the smallest eigenvalues~$s_k$ of the $S_k$ matrices for the exact case and the hardware results obtained with 8000 shots. The different columns correspond the different parameter sets shown in the panels of \autoref{fig:single_qubit_Sk_results}.}
  \label{tab:single_qubit_Sk_results}
\end{table}
\noindent differences. In general, the data from ibmq\_ourense (\autoref{fig:single_qubit_Sk_results}(a)), imbq\_santiago (\autoref{fig:single_qubit_Sk_results}(b)), ibmq\_valencia (\autoref{fig:single_qubit_Sk_results}(c)) and ibmq\_vigo (\autoref{fig:single_qubit_Sk_results}(d)) are in good agreement with the exact solution and, similar to our observations in the main text, do not show a strong dependence on the number of measurements. The shifts in the spectra of the $S_k$ matrices due to noise are also comparable for all the hardware backends we examine (see \autoref{tab:single_qubit_Sk_results}). While certain eigenvalues are slightly closer to the exact solution on some devices, none of them is clearly performing better than the others and we can confidently identify the parameter $\theta_4$ as the redundant one (see \autoref{tab:single_qubit_Sk_results}).

\subsection{EfficientSU2 for two qubits}\label{sec:hardware-implementation-2Q}
\begin{figure*}
  \begin{align*}
    \Qcircuit @C=1em @R=.7em {
      \lstick{\ket{0}} & \qw      & \gate{R_Y(\vartheta_1)} & \gate{R_Z(\vartheta_3)} & \ctrl{2} & \ctrl{1} & \gate{R_Y(\vartheta_5)} & \qw      & \gate{R_Z(\vartheta_7)} & \qw      & \qw    & \qw \\
      \lstick{\ket{0}} & \qw      & \gate{R_Y(\vartheta_2)} & \gate{R_Z(\vartheta_4)} & \qw      & \targ    & \gate{R_Y(\vartheta_6)} & \gate{Y} & \gate{R_Z(\vartheta_8)} & \qw      & \qw    & \qw \\
      \lstick{\ket{0}} & \gate{H} & \qw                     & \qw                     & \ctrl{0} & \qw      & \gate{X}                & \ctrl{-1} & \gate{X}                & \gate{H} & \meter & \cw \\
      }
  \end{align*}
  \caption{Circuit for measuring $\Re\langle\gamma_3,\gamma_6\rangle$ on quantum hardware. The ancillary qubit is initially put into superposition and acts as a control for the additional $CZ$ and $CY$ gates. A final measurement on the ancilla reveals the probability for obtaining $0$ which allows for obtaining $\Re\langle\gamma_3,\gamma_6\rangle$ according to \autoref{eq:real_part_overlap}.}
	\label{fig:Ak_circuit_two_qubits}
\end{figure*}
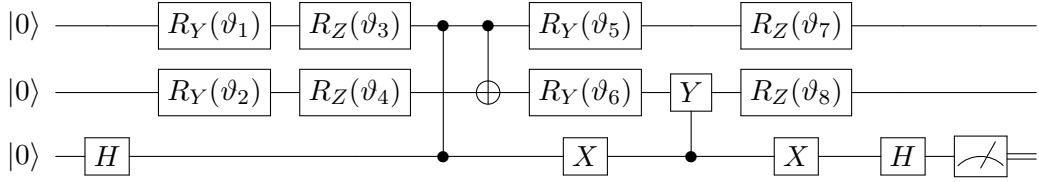

Our hardware efficient algorithm can also be readily applied to more complicated setups such as QISKIT's \verb|EfficientSU2| circuit. In the following we focus on the \verb|EfficientSU2| circuit for two-qubits and one repetition, $N=1$. This circuit has $8$ parameters $\theta_1,\dots,\theta_8$ originating from two blocks of $R_Y$ and $R_Z$ rotations, with a CNOT gate in between (see \autoref{sec:EfficientSU2}). Applying the analysis from \autoref{sec:EfficientSU2_again}, we find that the first $7$ parameters are independent and the final $R_Z(\theta_8)$ is superfluous. To check for that behavior with our algorithm, we proceed analogously to the single-qubit example. We draw again various random parameter sets, measure the entries of the $A_k$ vectors for those on quantum hardware and construct the $S_k$ matrices from these data. An example of the circuit used to obtain the entries of the vector $A_k$ is given in \autoref{fig:Ak_circuit_two_qubits}. Afterward we examine the spectra of the $S_k$ matrices using a classical computer. 

\begin{figure}[ht!]
	\centering
  \includegraphics[width=0.48\textwidth]{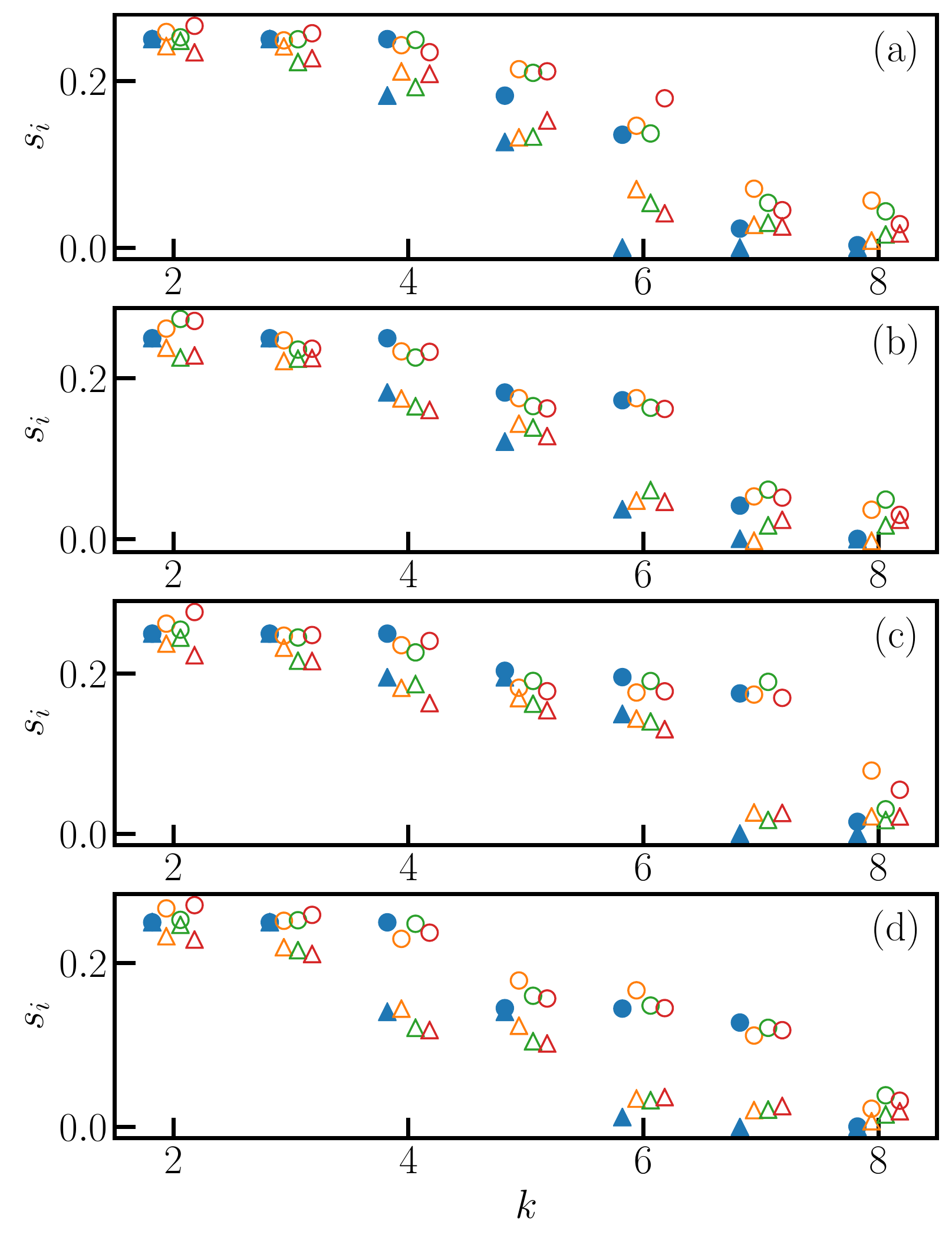}  
	\caption{Smallest (triangles) and second smallest (dots) eigenvalues of the matrices $S_k$, $k\geq 2$, for the \texttt{EfficientSU2} circuit for two qubits and a single repetition. The different panels each correspond to a randomly drawn parameter set $\theta_1,\dots,\theta_8$. The filled markers indicate the exact solution, the open markers the results obtained from evaluated on ibmq\_vigo with 1000 measurements (orange markers), 4000 measurements (gren markers) and 8000 measurements (red markers). The data points for each $k$ are slightly horizontally shifted for better visibility.}	\label{fig:two_qubit_Sk_results_ibmq_vigo}
\end{figure}

\begin{table}[ht!]
  \centering
  \begin{tabular}{|l||c|c|c|c|}
    \hline
    & (a) & (b) & (c) & (d)  \\ \hline \hline  
    $\frac{s_5}{10^{-4}}$ exact & 1267 & 1212 & 1955 & 1405\\ \hline
    $\frac{s_5}{10^{-4}}$ ibmq & 1525 & 1282 & 1544 & 1021\\ \hline \hline
    $\frac{s_6}{10^{-4}}$ exact & 1.632 & 371.5 & 1496 & 120.9\\ \hline
    $\frac{s_6}{10^{-4}}$ ibmq & 414.0 & 465.3 & 1307 & 366.1\\ \hline \hline
    $\frac{s_7}{10^{-4}}$ exact & 1.078 & 4.064 & 1.522 & 0.4057\\ \hline
    $\frac{s_7}{10^{-4}}$ ibmq & 256.1 & 242.1 & 263.7 & 257.2\\ \hline \hline
    $\frac{s_8}{10^{-4}}$ exact & 0 & 0 & 0 & 0\\ \hline
    $\frac{s_8}{10^{-4}}$ ibmq & 171.9 & 240.7 & 218.5 & 193.2\\ \hline 
  \end{tabular}  
  \caption{Numerical values for the smallest eigenvalues~$s_k$ of the $S_k$ matrices for the exact case and the hardware results obtained with 8000 shots. The different columns correspond the different parameter sets shown in the panels of \autoref{fig:two_qubit_Sk_results_ibmq_vigo}.}
  \label{tab:two_qubit_Sk_results_ibmq_vigo}
\end{table}

\begin{table}[ht!]
  \centering
  \begin{tabular}{|l||c|c|c|c|}
    \hline
    & (a) & (b) & (c) & (d)  \\ \hline \hline  
    $\frac{s_5}{10^{-4}}$ exact & 1955 & 1955 & 1955 & 1955\\ \hline
    $\frac{s_5}{10^{-4}}$ ibmq & 1949 & 1877 & 1553 & 1544\\ \hline \hline
    $\frac{s_6}{10^{-4}}$ exact & 1496 & 1496 & 1496 & 1496\\ \hline
    $\frac{s_6}{10^{-4}}$ ibmq & 1524 & 1729 & 1381 & 1307\\ \hline \hline
    $\frac{s_7}{10^{-4}}$ exact & 1.522 & 1.522 & 1.522 & 1.522\\ \hline
    $\frac{s_7}{10^{-4}}$ ibmq & 484.0 & 379.8 & 483.5 & 263.7\\ \hline \hline
    $\frac{s_8}{10^{-4}}$ exact & 0 & 0 & 0 & 0\\ \hline
    $\frac{s_8}{10^{-4}}$ ibmq & 449.2 & 218.2 & 450.7 & 218.5\\ \hline 
  \end{tabular}
  \caption{Numerical values for the smallest eigenvalues~$s_k$ of the $S_k$ matrices for the exact case and the hardware results obtained with 8000 shots. The different columns correspond the different parameter sets shown in the panels of \autoref{fig:two_qubit_Sk_results}.}
  \label{tab:two_qubit_Sk_results}
\end{table}  

\begin{figure}[ht!]
	\centering
  \includegraphics[width=0.48\textwidth]{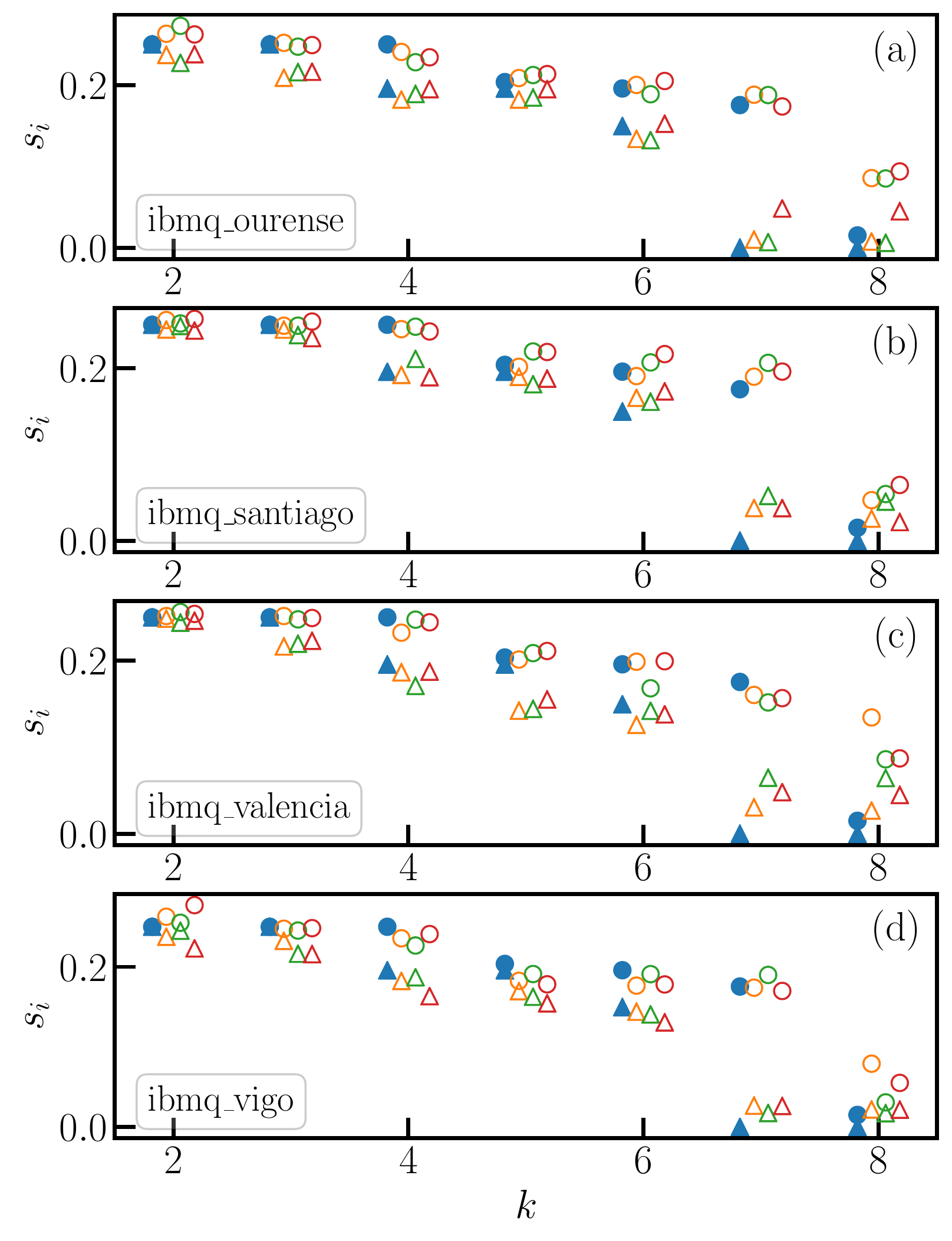}
	\caption{Smallest (triangles) and second smallest (dots) eigenvalues of the matrices $S_k$ for $k\geq 2$. The different panels correspond to a randomly drawn parameter set evaluated on the different chips (a) ibmq\_ourense, (b) ibmq\_santiago, (c) ibmq\_valencia, and (d) ibmq\_vigo. The filled markers indicate the exact solution, the open markers the results obtained from the quantum hardware with 1000 measurements (orange markers), 4000 measurements (green markers) and 8000 measurements (red markers). The data points for each $k$ are slightly horizontally shifted for better visibility.}
	\label{fig:two_qubit_Sk_results}
\end{figure}

Our results obtained on ibmq\_vigo for low-lying spectrum for these matrices are shown in \autoref{fig:two_qubit_Sk_results_ibmq_vigo} and \autoref{tab:two_qubit_Sk_results_ibmq_vigo}.
Focusing first on the exact results, we see that in this case it is harder to unambiguously identify the superfluous parameter in the circuit. While $S_8$ has as expected a vanishing eigenvalue for all the parameter combinations we study, \autoref{tab:two_qubit_Sk_results_ibmq_vigo} shows that the matrix $S_7$ has in general a very small, albeit non-vanishing eigenvalue of the order of $10^{-4} - 10^{-5}$. While we can easily resolve such a small difference for the exact results, the effects of noise on a quantum device can possibly obscure this small difference, thus making it hard to unambiguously detect the redundant gate.

Looking at the results obtained from ibmq\_vigo, we observe reasonable agreement with the exact solution. Compared to the single-qubit case, the dependence on the number of measurements is slightly stronger and, in general, the results deviate from the exact solution a little more (see \autoref{tab:two_qubit_Sk_results_ibmq_vigo}). The latter is to be expected because for the two-qubit case, the circuits necessary for computing the $S_k$ matrices are deeper and contain an additional CNOT gate compared to the single-qubit case. Consequently, our results are more hindered by noise.  In particular, \autoref{fig:two_qubit_Sk_results_ibmq_vigo}(b) and \autoref{fig:two_qubit_Sk_results_ibmq_vigo}(c) reveal that noise can indeed obscure the identification of the superfluous parameter. In those cases the smallest eigenvalues we obtain from our hardware data for $S_7$ and $S_8$ are quite small and very similar (see \autoref{tab:two_qubit_Sk_results_ibmq_vigo}), thus having the potential of misidentifying $\theta_7$ the interdependent parameter. The results for the parameter sets shown in \autoref{fig:two_qubit_Sk_results_ibmq_vigo}(a) and \autoref{fig:two_qubit_Sk_results_ibmq_vigo}(d) hint that $\theta_8$ is the superfluous parameter, however  the difference between the smallest eigenvalue of $S_7$ and $S_8$ is enhanced by noise in those cases. All in all, the data obtained from ibmq\_vigo to a certain degree still gives an indication that $\theta_8$ is interdependent, but compared to the single-qubit experiments the results are a lot less clear.

This picture also does not change using different hardware backends, as \autoref{fig:two_qubit_Sk_results} and \autoref{tab:two_qubit_Sk_results} demonstrate. Running the same randomly drawn parameter set on different quantum hardware, we observe rather similar performance. While the data from ibmq\_valencia in \autoref{fig:two_qubit_Sk_results}(c) seems slightly worse for the parameter set we study, none of the other backends can produce the exact solution with notably better accuracy than the others. Again, our results give a hint that the rotation gate associated with $\theta_8$ might be superfluous, but this is not unambiguously clear from the data, as \autoref{tab:two_qubit_Sk_results} shows.

Our results indicate that with the current noise levels in NISQ devices, it might not be possible to definitely identify the independent parameters in a circuit. Nevertheless, as we approach the interdependent parameters, there is still some signal in the data that might allow to get an idea which parameter is superfluous.

\bibliographystyle{plainnat}
\bibliography{Papers}

\end{document}